\documentclass[aps,prx,preprint,onecolumn,citeautoscript,footinbib,
eqsecnum]{revtex4-1}  
\usepackage{amsfonts,amsthm,amsmath,amssymb,upgreek}
\usepackage{amsmath,amssymb,bm} 
\usepackage{graphicx}  
\usepackage[export]{adjustbox}
\usepackage{color} 
\usepackage[dvipsnames]{xcolor}
\usepackage[papersize={8.5in,11in}]{geometry}
\usepackage[colorlinks=true]{hyperref}  
\usepackage{ulem}
\usepackage[mathscr]{euscript}
\usepackage[section]{placeins}
\hypersetup{
    bookmarks=true,         
    unicode=false,          
    pdftoolbar=true,        
    pdfmenubar=true,        
    pdffitwindow=false,     
    pdfstartview={FitH},    
    pdfsubject={},   
    pdfcreator={},   
    pdfproducer={}, 
    pdfkeywords={} {} {}, 
    pdfnewwindow=true,      
    colorlinks=true,       
    linkcolor=magenta, 
    citecolor=blue,        
    filecolor=magenta,      
    urlcolor=blue           
} 

\usepackage{subfig}

\usepackage{amsfonts}
\usepackage{upgreek}
\usepackage{slashed}
\usepackage{latexsym}

\newcommand{\beq}{\begin{equation}}
\newcommand{\eeq}{\end{equation}}
\def\bea{\begin{eqnarray}}
\def\eea{\end{eqnarray}}
\newcommand{\nn}{\nonumber \\}
\renewcommand{\vec}[1]{\boldsymbol{#1}}

\makeatletter
\newcommand{\mathleft}{\@fleqntrue\@mathmargin0pt}
\newcommand{\mathcenter}{\@fleqnfalse}
\makeatother

\begin{document}


\title{Metal-insulator transition in a random Hubbard model}

\author{Grigory Tarnopolsky}
\affiliation{Department of Physics, Harvard University, Cambridge MA 02138, USA}

\author{Chenyuan Li}
\affiliation{Department of Physics, Harvard University, Cambridge MA 02138, USA}

\author{Darshan G. Joshi}
\affiliation{Department of Physics, Harvard University, Cambridge MA 02138, USA}

\author{Subir Sachdev}
\affiliation{Department of Physics, Harvard University, Cambridge MA 02138, USA}

\date{\today
\\
\vspace{0.4in}}

\begin{abstract}
We examine the metal-insulator transition in a half-filled Hubbard model of electrons with random and all-to-all hopping and exchange, and an on-site non-random repulsion, the Hubbard $U$. We argue that recent numerical results of Cha {\it et al.\/} (\href{https://arxiv.org/abs/2002.07181}{arXiv:2002.07181}) can be understood in terms of a deconfined critical point between a disordered Fermi liquid and an insulating spin glass. We find a deconfined critical point in a previously proposed large $M$ theory which generalizes the SU(2) spin symmetry to SU($M$), and obtain exponents for the electron and spin correlators which agree with those of Cha {\it et al.\/}. We also present a renormalization group analysis, and argue for the presence of an additional metallic spin glass phase at half-filling and small $U$.
\end{abstract}

\maketitle
\tableofcontents

\section{Introduction}
\label{sec:intro}

The Mott metal-insulator transition is central to an understanding of correlated electrons \cite{Imada98}. 
In many three-dimensional correlated electron compounds, and in dynamic mean-field theories, this transition is first order. However, there are cases when the transition can be continuous, with interesting possibilities for non-Fermi liquid and `strange metal' behavior at non-zero temperature in the vicinity of the critical point. One case which has been much studied theoretically \cite{Hermele07,Senthil08,Podolsky09,Krempa12} is when the Mott insulator is a spin liquid with a spinon Fermi surface, and the continuous transition involves condensation of an electrically charged boson which also carries charges under an emergent gauge field. 

\begin{figure}[tb]
\begin{center}
\includegraphics[height=3.2in]{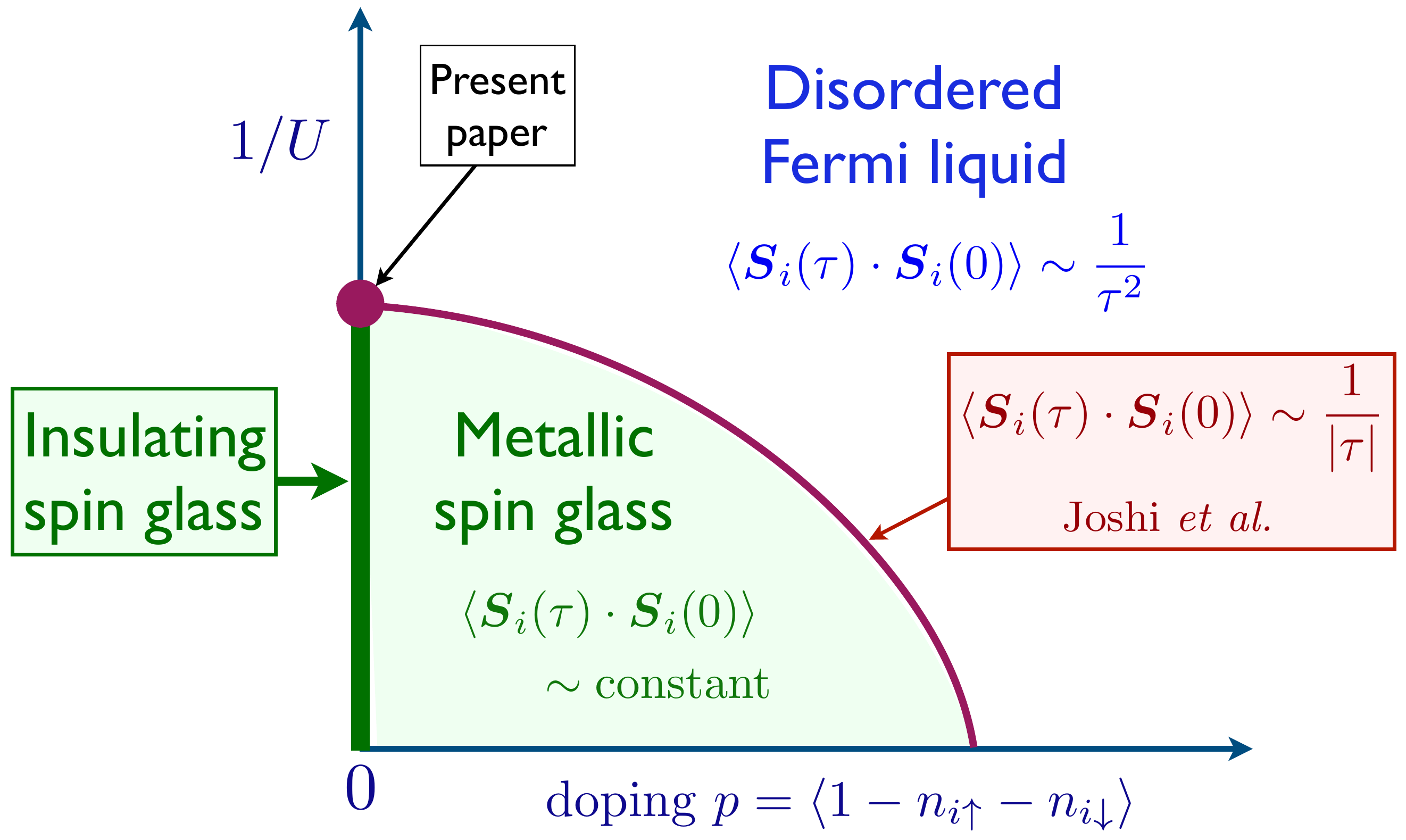} 
\end{center}
\caption{Proposed phase diagram of the random Hubbard model in (\ref{HU}) with SU(2) spin symmetry. The present paper describes the metal insulator transition at $p=0$ between the insulating spin glass and the disordered Fermi liquid as a deconfined critical point in Section~\ref{sec:largeM}. The non-zero $p$ transition between two metallic states at large $U$ was described by Joshi {\it et al.\/} \cite{Joshi2019} as deconfined critical point in a $t$-$J$ model.}
\label{fig:phasediag}
\end{figure}

In the present paper, we will focus on the continuous (or nearly continuous) Mott transition observed recently in a numerical study of a Hubbard model supplemented by random exchange interactions by Cha {\it et al.\/} \cite{Cha19}. Such a model was previously studied by Florens {\it et al.\/} \cite{Florens13} using a large $M$ approach which generalized the SU(2) spin symmetry to SU($M$). In the large $M$ limit, the saddle point equations obtained by Florens {\it et al.\/} \cite{Florens13} turn out to be essentially identical to the saddle point equations of a different model studied recently by Fu {\it et al.} \cite{Fu2018}. Fu {\it et al.} \cite{Fu2018} obtained analytic results on the low energy structure of gapless states in their model, and so we can transfer their results to the random Hubbard model of Florens {\it et al.\/} \cite{Florens13} and Cha {\it et al.\/} \cite{Cha19}. We will find that the large $M$ exponents  obtained by Fu {\it et al.} \cite{Fu2018} for the critical state in Section~\ref{sec:delta14} agree with the corresponding exponents for the electron and spin correlators at the continuous Mott transition obtained numerically Cha {\it et al.} \cite{Cha19} for the case with SU(2) spin symmetry.
In the large $M$ theory, the Mott criticality is realized by a deconfined critical point \cite{senthil1}, described by the 
fractionalization of the electron into fermionic spinons and charged scalars both carrying an emergent U(1) gauge charge. We argue that this deconfined critical point separates a disordered Fermi liquid from an insulating spin glass (see Fig.~\ref{fig:phasediag}).

In Section~\ref{sec:RG}, we will present a renormalization group (RG) study of the random Hubbard model for the case with SU(2) spin symmetry (and also for general SU($M$)). As in a recent study of the random $t$-$J$ model \cite{Joshi2019}, the RG is performed on a quantum impurity model, with the impurity site coupled to fermionic and bosonic baths, and supplemented by  self-consistency conditions. The RG analysis follows methods developed in Refs.~\onlinecite{VBS2000,FritzVojta04}. For the particle-hole symmetric case relevant to half-filling, the RG requires a perturbative treatment of 
the on-site repulsion (the Hubbard $U$) between the electrons in the context of an $\epsilon$ expansion (defined in (\ref{QRpower})). The solution of the self-consistency conditions requires an extrapolation to $\epsilon=1$; unlike the previous work \cite{Joshi2019}, we are unable to perform this extrapolation with any reliability as we do not have access to the needed exponents to all orders in $\epsilon$. 

The RG analysis in the small $U$,$\epsilon$ expansion yields a finite coupling fixed point with one relevant direction. This fixed point is a candidate to describe the metal-insulator transition at $p=0$ in Fig.~\ref{fig:phasediag}, with the larger $U$ direction away from the fixed point flowing to the insulating spin glass state. However, we don't really have control over the computation far from the fixed point, and it is possible that the fixed point actually describes the onset of metallic spin glass order from the disordered Fermi liquid, as indicated in the phase diagram in Fig.~\ref{fig:phasediag2}. 
We also note that there is a previous Landau-type theory \cite{SRO1995,Sengupta95} for such a metal-metal transition, and this will be reviewed in the  present small $U$ context in Appendix~\ref{app:metglass}.

We turn to a description of the model of interest in this paper, for the case with SU(2) spin symmetry.
We consider electrons, annihilated by $c_{i \alpha}$, spin $\alpha = \uparrow, \downarrow$ on $N$ sites $i=1 \ldots N$ with the Hamiltonian
\beq
H = \sum_{i = i}^N \left( -\mu (n_{i\uparrow} + n_{i \downarrow}) + U n_{i \uparrow} n_{i \downarrow} \right) + \frac{1}{\sqrt{N}} \sum_{i \neq j=1}^N t_{ij}
 c_{i\alpha}^\dagger c_{j \alpha}^{}  + \frac{1}{\sqrt{N}} \sum_{i < j=1}^N J_{ij} \vec{S}_i \cdot \vec{S}_j \label{HU}
\eeq
where $\mu$ is the chemical potential, 
\beq
n_{i \alpha} = c_{i \alpha}^\dagger c_{i \alpha}^{} \quad , \quad 
\vec{S}_i = \frac{1}{2} c_{i \alpha}^\dagger \vec{\sigma}_{\alpha\beta} c_{i \beta}^{}
\eeq
are the number and spin operators with $\vec{\sigma}$ the Pauli matrices. The density of the electrons is specified by the filling $p$
\beq
p = \left \langle 1 - n_{i\uparrow} - n_{i \downarrow} \right \rangle\,.
\eeq
We can take the $t_{ij}$ to be all equal between the sites of a Bethe lattice with large co-ordination number, or use a fully connected cluster in which all $t_{ij}= t_{ji}^\ast$ are independent random variables with zero mean and $\overline{ |t_{ij}|^2} = t^2$. We will focus on the random case because it is a bit simpler, but equivalent results apply to the Bethe lattice. 
The real exchange interactions $J_{ij}$ are independent random numbers with zero mean and mean-square value $\overline{J_{ij}^2} = J^2$. 

Let us take a broader perspective, and consider the phase diagram of $H$ as a function of $U$ and hole density away from half-filling, $p=0$; see Fig.~\ref{fig:phasediag}.
At large $U$ and $p=0$, we have an insulating spin glass state: at $p=0$ we need only consider the spin-only model with the $J_{ij}$ interactions, and a spin glass state was found in numerical studies \cite{MJR02,MJR03}, in contrast to the critical spin liquid appearing in the large $M$ limit \cite{SY92}. We will be interested here in the approach to the spin glass insulator at $p=0$ from the small $U$ side, across a metal-to-insulator transition from a disordered Fermi liquid at small $U$ and $p=0$.
Upon doping the spin glass, we expect a metallic spin glass state for a range of non-zero $p$, before there is a distinct quantum phase transition to a disordered Fermi liquid state at a nonzero $p$: this large $U$ transition is also expected to be described by a deconfined critical point, and is discussed in a separate paper \cite{Joshi2019}.

The outline of the paper is as follows. Section~\ref{sec:largeN} will described the limit of a large number of sites, $N$, where $H$ is mapped onto a non-local in time effective action for a single site with self-consistency conditions on its correlators. Section~\ref{sec:largeM} will describe the solution of the single site problem for the case where the SU(2) spin symmetry is generalized to SU($M$) with $M$ large: we will describe the correspondence between the large $M$ solutions, and the numerical results of 
Cha {\it et al.\/} \cite{Cha19} for $M=2$. Section~\ref{sec:RG} presents the RG analysis of the single site model with SU(2) symmetry obtained in Section~\ref{sec:largeN}. Appendix~\ref{app:metglass} reviews the theory of Ref.~\onlinecite{SRO1995} for the onset of metallic spin glass order in a disordered Fermi liquid in a conventional Landau-type transition (and not via a large $U$ deconfined critical point \cite{Joshi2019}) which can be present at small $U$, as indicated in Fig.~\ref{fig:phasediag2}.

\section{Large volume limit}
\label{sec:largeN}

The limit of large volume ($N \rightarrow \infty$) of $H$ is obtained by the methods described in Refs.~\cite{SY92,GPS00,GPS01,Cha19,Joshi2019}.
We introduce field replicas in the path integral, and average over $t_{ij}$ and $J_{ij}$. 
At the $N=\infty$ saddle point, the problem reduces to a single site
problem, with the fields carrying replica indices. 
The replica structure is important in the spin glass phase \cite{GPS00,GPS01}. 
In the interests of simplicity, we drop the replica indices here as they play no significant role in the critical theory and the RG equations. 
Within the imaginary time path integral formalism (with $\tau\in[0,1/T]$, with $T$ the temperature), 
the solution of the model involves a local single-site effective action which reads:
\begin{eqnarray}
\mathcal{Z} &=& \int \mathcal{D} c_\alpha (\tau) e^{-\mathcal{S}} \nonumber \\
\mathcal{S} &=& \int d \tau \left[c_\alpha^\dagger (\tau)  \left(\frac{\partial}{\partial \tau}-\mu \right)  c_\alpha (\tau) + \frac{U}{2} c^\dagger_\alpha (\tau) c^\dagger_{\beta} (\tau) c_{\beta}^{} (\tau) c_{\alpha}^{} (\tau) \right] \nonumber \\
&~&~~ - t^2 \int d\tau d \tau' R (\tau - \tau') c_\alpha^\dagger (\tau) c_\alpha (\tau')  - \frac{J^2}{2} \int d\tau d \tau' Q (\tau - \tau') \vec{S} (\tau) \cdot \vec{S} (\tau') \,, \label{Z}
\end{eqnarray}
In this expression, $\mu$ is the chemical potential chosen to ensure $p=0$.
Decoupling the path integral introduces fields analogous to $R$ and $Q$ which are initially off-diagonal in the spin SU(2) indices. 
We have assumed above that the large-volume  limit is dominated by the saddle point in which
spin rotation symmetry is preserved on the average, and so $R$ and $Q$ were taken to diagonal in spin indices.
The path integral $\mathcal{Z}$ is a functional of the fields $R(\tau)$ and $Q(\tau)$, and we define its correlators 
\begin{eqnarray}
\overline{R}(\tau - \tau') &=& -  \left\langle c^{}_{\alpha} (\tau) c^{\dagger}_\alpha (\tau') \right\rangle_\mathcal{Z} \nonumber \\
\overline{Q} (\tau - \tau') &=&  \frac{1}{3} \left\langle \vec{S} (\tau) \cdot \vec{S} (\tau') \right\rangle_\mathcal{Z} 
\label{self0}
\end{eqnarray}
In the thermodynamic ($N\rightarrow\infty$) limit, 
the solution of the model is obtained by imposing the two self-consistency conditions:  
\beq
R(\tau) = \overline{R}(\tau) \quad, \quad Q(\tau) = \overline{Q} (\tau).
\label{selfcons}
\eeq
These equations and the mapping to a local effective action are part of the extended dynamical mean-field theory framework (EDMFT), which becomes 
exact for random models on fully connected lattices~\cite{Sengupta95}.
They can also be viewed as an EDMFT approximation to non-random 
models~\cite{Smith2000,Haule02,Haule03,Haule07}. 

\section{Large $M$ theory}
\label{sec:largeM}
 
We consider here a large $M$ generalization of the $N$-site Hubbard model, following
Refs.~\cite{Florens02,Florens04,Florens13,Fu2018}, and examine the structure of the large $M$ limit at $N=\infty$. 
We consider an electron $c_{p,\alpha}$ with a spin index $\alpha = 1 \ldots M$, and an `orbital' index $p = 1 \ldots M'$. We will take
the limit of large number of sites, $N$, followed by the limit of large $M$ and $M'$
at fixed 
\beq
k \equiv \frac{M'}{M} \,.
\eeq
We are interested in the case with SU(2) spin symmetry which has the values $M=2$, $M'=1$, $k=1/2$.
The large $M$,$M'$ limit requires us to fractionalize the electron as
\beq
c_{i p \alpha}^\dagger = X_{ip} f_{i \alpha}^\dagger \,, \label{defc}
\eeq
where $X_{ip}$ is a complex `slave rotor' \cite{Florens02,Florens04}, with $p=1 \ldots M'$, obeying the constraint
\beq
\sum_{p=1}^{M'} |X_{ip}|^2 = M' \,.
\label{unit}
\eeq
This representation has a U(1) gauge invariance
\beq
X_{ip} \rightarrow X_{ip} e^{i \phi_i (\tau)} \quad , \quad f_{i \alpha} \rightarrow f_{i \alpha} e^{i \phi_i (\tau)} \label{u1gauge}
\eeq
We shall be interested in the sector in which the U(1) gauge charge is fixed on each site by \cite{Florens02}
\beq
\sum_{\alpha=1}^M f_{i \alpha}^\dagger f_{i \alpha} + \hat{L}_i = \frac{M}{2}
\label{const}
\eeq
where $\hat{L}_i$ is the U(1) angular momentum operator for the rotors.

The Hamiltonian generalizing (\ref{HU}) we shall study in this section is a combination of those in Refs.~\cite{SY92,Florens02,Florens13}:
\bea
H &=& \frac{U}{2 M'} \sum_{i} \left( \sum_{\alpha=1}^M f_{i \alpha}^\dagger f_{i \alpha} - \frac{M}{2}  \right)^2 + \epsilon_0 \sum_{i p\alpha} f_{i \alpha}^\dagger f_{i \alpha} \nonumber \\
&~&~~~+ \frac{1}{\sqrt{NM}}\sum_{i,j,p,\alpha} t_{ij} c_{i p \alpha}^\dagger c_{j p \alpha} +  \frac{1}{\sqrt{NM}}\sum_{i>j,\alpha\beta} J_{ij} f_{i \alpha}^\dagger f_{i \beta} f_{j \beta}^\dagger f_{j \alpha} \,.
\label{Hrotor}
\eea
The value of $\epsilon_0$ is adjusted to fix the average electron density at each site, $\sum_{\alpha=1}^M f_{i \alpha}^\dagger f_{i \alpha}$ to equal $M/2$ for the half-filled case. 

We now take the large volume limit of Section~\ref{sec:largeN}. For (\ref{Hrotor}), the $N \rightarrow \infty$ 
limit reduces to the following single-site path integral (replacing (\ref{Z}))
\bea
\mathcal{Z} &=& \int \mathcal{D} f_\alpha \mathcal{D} X_p \mathcal{D} \lambda \mathcal{D} h
e^{- \mathcal{S}} \nn 
\mathcal{S} &=& \int_0^{1/T} d \tau \left[ \frac{1}{2U}\sum_{p}\left| \left(\frac{\partial}{\partial \tau} + i h \right) X_{p} \right|^2 
+ i \lambda \left(\sum_p |X_{p}|^2 - M' \right)
+ \sum_{\alpha} f_{\alpha}^\dagger \left( \frac{\partial}{\partial \tau} + \epsilon_0 + i h \right) f_{\alpha} - i h \frac{M}{2} \right] \nn 
&~&~-\frac{t^2}{M} \sum_{p,\alpha} \int_{0}^{1/T} d \tau d \tau' R^\ast (\tau - \tau') 
X_{p} (\tau) X^\ast_{p} (\tau') f_{\alpha}^\dagger (\tau ) 
f_{\alpha} (\tau ')  \nn  &~&~-
\frac{J^2}{2M} \sum_{\alpha,\beta} \int_{0}^{1/T} d \tau d \tau'  Q(\tau-\tau') f_{\alpha}^\dagger (\tau) f_{\beta} (\tau) f_{\beta}^\dagger (\tau') f_{\alpha} (\tau') \,.
\label{L}
\eea
Here $T$ is the temperature, $\lambda$ is the Lagrange multiplier imposing Eq.~(\ref{unit}) and $h$ is the Lagrange multiplier imposing Eq.~(\ref{const}). The U(1) gauge invariance
(\ref{u1gauge}) applies also to (\ref{L}) after we transform
\beq
h \rightarrow h - \partial_\tau \phi\,.
\eeq
The self-consistency equations (\ref{selfcons}) now become
\bea
R(\tau - \tau') &=& -\frac{1}{M M'} \sum_{p,\alpha} \left\langle X_{p} (\tau) X^\ast_{p} (\tau') f_{\alpha}^\dagger (\tau ) 
f_{\alpha} (\tau ') \right\rangle_\mathcal{Z} \nn 
Q(\tau - \tau') &=& \frac{1}{M^2} \sum_{\alpha,\beta} \left\langle f_{\alpha}^\dagger (\tau) f_{\beta} (\tau) f_{\beta}^\dagger (\tau') f_{\alpha} (\tau') \right\rangle_\mathcal{Z} \,.
\label{saddle}
\eea

Having taken the large $N$ limit, we can now take the large $M$, $M'$ limit at fixed $k=M'/M$. 
We have set things up so we can decouple the quartic terms in $\mathcal{S}$ by Hubbard-Stratonovich fields, perform the path integrals over $f_\alpha$ and $X_p$, 
and then perform a $1/M$ expansion about the saddle point.
The large $M$ saddle point equations are essentially those obtained in Ref.~\onlinecite{Fu2018}, and we adapt the relevant analysis here. We limit ourselves to the particle-hole symmetric case with $p=0$, in which case we can set $\epsilon_0 =0$ and $h=0$.
We obtain for the $f$ fermion Green's function, $G_f$, and the $X$ correlator $\chi$
\bea
G_f (i \omega_n) &=& \frac{1}{i \omega_n - \Sigma_f (i \omega_n)} \quad, \quad \Sigma_f (\tau) = - J^2 G_f^2 (\tau) G_f (-\tau)
+ k \, t^2 G_f (\tau) \chi^2 (\tau) \label{feqns} \\
\chi ( i \omega_n) &=& \frac{1}{\omega_n^2 /U+ \chi_0^{-1} -  P (i \omega_n) + P(i\omega_n=0)} \quad, \quad P (\tau) = - t^2 G_f (\tau) G_f (-\tau) \chi (\tau) \label{beqns}
\eea
where 
\beq 
i\lambda = \chi_0^{-1} + P(i \omega_n = 0)
\label{lambda0}
\eeq
is the saddle point value of $i \lambda$. 
Note that we have introduced notation so that
\beq
\chi (i\omega_n = 0) \equiv \chi_0 \,,
\label{static}
\eeq
is the static $X$ susceptibility. Formally, the value of $\chi_0$ is to be determined by solving the constraint equation Eq.~(\ref{unit}):
\beq
T \sum_{\omega_n} \chi (i \omega_n) = 1\,. \label{gval}
\eeq

\begin{figure}
\begin{center}
\includegraphics[height=3.2in]{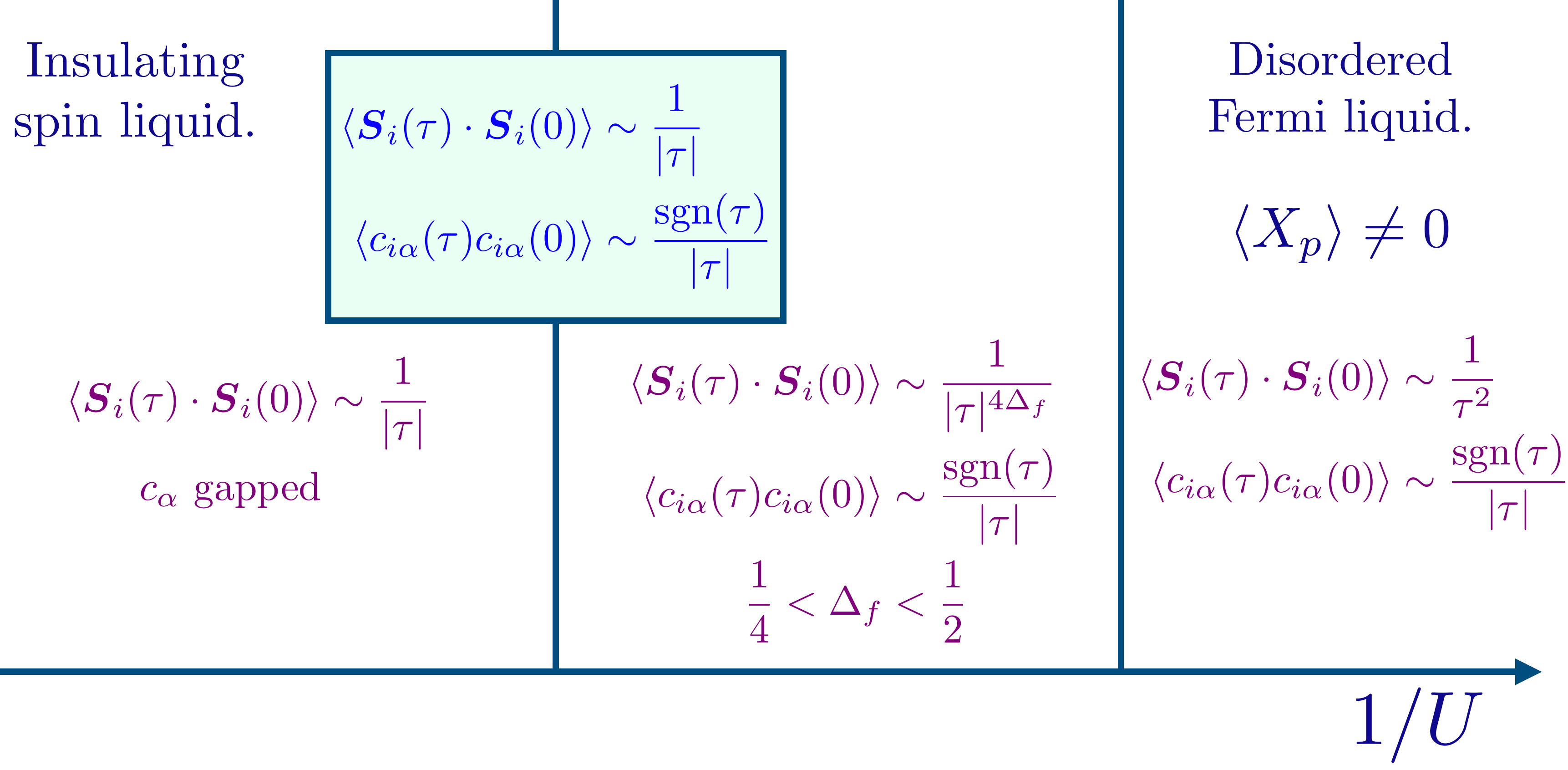} 
\end{center}
\caption{Schematic phase diagram in the large $M$ limit. This section does not describe the disordered Fermi liquid where the $X_p$ condense; for $M'>1$, this phase also has orbital glass order, associated with the orbital index $p=1 \ldots M'$. For the physical case $M=2$, $M'=1$, the insulating spin liquid is expected to be replaced by an insulating spin glass \cite{MJR02,MJR03}, and we argue that the intermediate critical phase with $1/4 < \Delta_f < 1/2$ may not exist.}
\label{fig:phasediag:M}
\end{figure}

The saddle-point equations (\ref{feqns}), (\ref{beqns}), and (\ref{gval}) were examined numerically and analytically in Ref.~\onlinecite{Fu2018} in the context of a different model. Here we transfer their analysis to our model. We recall the analytic low energy structure of the solutions, starting with large $U$, and then with decreasing $U$. At very large $U$, we expect the $X$ boson correlators to decay rapidly in time, and to become progressively longer ranged as $U$ is decreased.
A sketch of our proposed large $M$ phase diagram is shown in Fig.~\ref{fig:phasediag:M}.

\subsection{Gapped boson}
\label{sec:gappedboson}

At very large $U$, we expect an energy gap in the boson correlator $\chi (\omega)$, corresponding to exponential decay of $X$ correlators, and the Mott gap in an insulator. In this case, we can simply drop the boson Green's function in (\ref{feqns}) at low energy, and (\ref{feqns}) reduces to the equations of the spin-only model examined originally in Ref.~\onlinecite{SY92}. As argued there, the fermion Green's function is gapless with the large imaginary time ($\tau \rightarrow \infty$) form at $T=0$
\beq
G_f (\tau) = - \mbox{sgn} (\tau) \frac{F}{|\tau|^{2 \Delta_f}}\,. \label{Gcrit}
\eeq
The exponent $\Delta_f = 1/4$ \cite{SY92}. Although the $f$ fermion is gapless, the $X$ boson is gapped, and so the electron $c$ is also gapped, and this solution describes an insulator. The spin correlations in this insulator are however gapless: the spin operator is
$S^a = c_{p \alpha}^\dagger T^a_{\alpha\beta} c_{p \beta}$, where $T^a$, with $a=1\ldots M^2-1$, is a generator of SU($M$), and it has the long-time correlator
\beq
\left\langle S^a (\tau) S^a (0) \right\rangle \sim \frac{1}{|\tau|^{4 \Delta_f}} \label{Sc} \,.
\eeq
With $\Delta_f=1/4$, we conclude that the spin correlator decays as $1/|\tau|$. From numerical studies of the insulating quantum magnet \cite{MJR02,MJR03}, we now know that the present insulating, gapless `spin-fluid' solution is present only at large $M$. For the case $M=2$ of interest to us, the insulator has spin-glass order. So the gapped $X$ solution discussed here should be mapped to the insulating spin glass state found by Cha {\it et al.\/} \cite{Cha19} at large $U$.

The nature of the gapped boson correlator was also examined in Ref.~\onlinecite{Fu2018}, and it was found that 
\beq
\chi (\tau) \sim \frac{e^{-m |\tau|}}{\sqrt{|\tau|}}
\eeq
at large $|\tau|$, where $m$ is the Mott gap.

\subsection{Gapless boson}
\label{sec:gaplessboson}

With decreasing $U$, we expect solutions in which the boson $X$ is critical or condensed, as shown in Fig.~\ref{fig:phasediag:M}. We consider the critical case.
Along with the low frequency form for the fermion in (\ref{Gcrit}), we assume a power-law form for the boson correlator at long times $\tau$:
\beq
 \chi (\tau) = \frac{C}{|\tau|^{2 \Delta_b}} \,. \label{chicrit}
\eeq
We will find below that consistency requires that $\Delta_b \leq 1/4$, and so from (\ref{static}) $\chi_0 = \infty$ due to a IR divergence.
In the large $M$ limit, the ansatzes (\ref{Gcrit}) and (\ref{chicrit}) imply that the gauge-invariant electron Green's function decays as
\beq
G_c (\tau) = - \left\langle c_{p\alpha} (\tau) c_{p\alpha}^\dagger (0) \right\rangle \sim -\frac{\mbox{sgn}(\tau)}{|\tau|^{2 (\Delta_f + \Delta_b)}} \,.
\label{Gc}
\eeq

Following Ref.~\onlinecite{Fu2018}, we will now see by explicit computation that the ansatzes (\ref{Gcrit}) and (\ref{chicrit}) are indeed valid solutions of the saddle point equations (\ref{feqns}) and (\ref{beqns}) at long times. 
First, we need
the Fourier transforms at $T=0$ which are at small $\omega$
\bea
G_f (i \omega) &=& - 2i \, \mbox{sgn} (\omega)\frac{F}{|\omega|^{1 - 2 \Delta_f}} \cos (\pi \Delta_f) \Gamma( 1- 2 \Delta_f) \nn 
\chi (i \omega) &=&  2 \frac{C}{|\omega|^{1 - 2 \Delta_b}} \sin (\pi \Delta_b) \Gamma( 1- 2 \Delta_b)\,.
\label{g2}
\eea
From Eq.~(\ref{feqns}) and (\ref{beqns}), the self energies are 
\bea
\Sigma_f (\tau ) &=& - \mbox{sgn} (\tau) \left(\frac{J^2 F^3}{|\tau|^{6 \Delta_f}} + \frac{ k \,t^2 F C^2}{|\tau|^{2 \Delta_f + 4 \Delta_b}} \right) \nn
P(\tau) &=&  \frac{t^2 F^2 C}{|\tau|^{4 \Delta_f + 2 \Delta_b}}\,,
\eea
and their Fourier transforms are 
\bea
\Sigma_f (i \omega ) &=& - 2 i \, \mbox{sgn} (\omega) \left(\frac{J^2 F^3}{|\omega|^{1-6 \Delta_f}}
\cos(3 \pi \Delta_f) \Gamma(1 - 6 \Delta_f) \right. \nn 
&~&~~~~~~~~\left. 
 + \frac{ k  t^{2} F C^2}{|\omega|^{1-2 \Delta_f - 4 \Delta_b}} 
 \cos(\pi (\Delta_f + 2 \Delta_b)) \Gamma(1 - 2 \Delta_f - 4 \Delta_b)
 \right) \nn 
P(i \omega) - P(i\omega=0) &=&  2 \frac{ t^{2} F^2 C}{|\omega|^{1-4 \Delta_f - 2 \Delta_b}}
\sin(\pi (2 \Delta_f + \Delta_b)) \Gamma(1 - 4\Delta_f - 2 \Delta_b)
\,. \label{se2}
\eea
From Eqns~(\ref{g2}) and (\ref{se2}), and using $G_f (i \omega) \Sigma_f  (i \omega) = -1$
and $\chi (i \omega) (P(i \omega)-P(i\omega=0)) = -1$ in the limit of low $\omega$, we see that solutions are only possible
when
\beq
\Delta_f + \Delta_b = 1/2 \,. \label{Deltafb}
\eeq
From (\ref{Gc}) we now see that the electron Green's function $G_c (\tau) $ always has the decay $\sim 1/\tau$, which is the same as that in a Fermi liquid. This is in agreement with the electron correlator obtained by Cha {\it et al.} \cite{Cha19} at the metal-insulator critical point, and in the Fermi liquid phase.

Further examination of the saddle point equations shows that two classes of solutions are possible,  depending upon whether $\Delta_f > 1/4$ 
or $\Delta_f = 1/4$. We will examine these solutions in the following subsections.

\subsubsection{$\Delta_f = \Delta_b =  1/4$}
\label{sec:delta14}

In this case, both terms in $\Sigma$ in Eq.~(\ref{se2}) have the same low frequency power-law, and so both contribute to the low $\omega$ limit.
The Schwinger-Dyson equations have solutions which reduce to
\bea
J^2  F^4 + k t^2 C^2 F^2 &=& \frac{1}{4 \pi} \nn 
 t^2 C^2 F^2 &=& \frac{1}{4 \pi} \,. \label{e2}
\eea
These can be solved uniquely for both $F>0$ and $C>0$ provided again $k < 1$. 
The existence of a unique low $\omega$ solution with these exponents indicates that Eq.~(\ref{gval}) will be satisfied at only a particular
value of the couplings {\it i.e.\/} this solution corresponds to a critical point as $U$ is decreased to smaller values from the gapped boson phase: numerical evidence for this structure was obtained by Fu {\it et.al.\/} \cite{Fu2018}.

Consequently, we identify the present $\Delta_f = \Delta_b = 1/4$ solution with the metal-insulator critical point found by Cha {\it et al.} \cite{Cha19}. Indeed, via (\ref{Sc}) the spin correlator decays as $1/|\tau|$, and via (\ref{Gc}), the electron correlator decays as $1/\tau$, and these correspond to the leading exponents found numerically by Cha {\it et al.} \cite{Cha19}.

Although the $1/\tau$ decay of the electron correlator is the same as that of a Fermi liquid, the spin correlator is distinct from the $1/\tau^2$ decay in a Fermi liquid. Indeed, the presence of a $1/\tau$ electron correlator {\it and} a $1/|\tau|$ spin correlator is evidence for fractionalization at this critical point: both correlators are simply understood from a fractionalization of $c$ into $X$ and $f$ in (\ref{defc}), and from the scaling dimensions $\Delta_f = \Delta_b =1/4$.

\subsubsection{$\Delta_f > 1/4$}
\label{sec:deltag14}

With a further decrease in $U$, Ref.~\onlinecite{Fu2018} found that we should consider the case with a faster decrease in spin correlations.
With $\Delta_f > 1/4$, the first term in $\Sigma_f (i \omega)$ in Eq.~(\ref{se2}) is subdominant and can be ignored.
Then the Schwinger-Dyson equations can be solved, and they simplify to the relations
\bea
 k t^2 F^2 C^2 \frac{4 \pi \cot(\pi \Delta_f)}{2 - 4 \Delta_f} &=& 1 \nn 
 t^2 F^2 C^2  \frac{\pi \cot(\pi \Delta_f)}{\Delta_f}&=& 1\,.
\label{eq:FC}
\eea
Note that these equations are independent of $J$, and so the asymptotic low energy structure does not depend upon the strength of the exchange interactions.
They are consistent only if we choose the scaling dimensions
\beq
\Delta_f = \frac{1}{2k + 2} \quad , \quad \Delta_b = \frac{k}{2 k + 2} \,.
\label{alphaexp}
\eeq
Note that $\Delta_f > 1/4$ requires $k < 1$. So the exponents are limited to the ranges
\beq
\frac{1}{4} < \Delta_f < \frac{1}{2} \quad , \quad 0 < \Delta_b < \frac{1}{4} \,.
\eeq
This analysis of the low $\omega$ limit of the saddle point equations does not determine the values of $F$ and $C$ separately, only
the value of their product $C F$. So we expect that the $\Delta_f > 1/4$ solution defines a critical phase which extends over a range of value of the couplings. 

Ref.~\onlinecite{Fu2018} labeled this critical phase as `quasi-Higgs'. The numerical results of Cha {\it et al.} \cite{Cha19} do not indicate such an extended critical phase for the SU(2) case. It is possible that such a phase only appears for larger $M$, and is absent, or very small in extent, for $M=2$. 

We now argue that with a further decrease in $U$, the quasi-Higgs phase will be replaced by an actual Higgs phase, as sketched in Fig.~\ref{fig:phasediag:M}. Note from (\ref{se2}) and (\ref{Deltafb}) that $P(i \omega) - P(0) \sim |\omega|^{2 \Delta_f}$, and so with $\Delta_f < 1/2$, the frequency integral in (\ref{gval}) has no infra-red divergence (recall $\chi_0^{-1} = 0$). Consequently \cite{GPS00,GPS01}, at small enough $U$ we will not be able to satisfy (\ref{gval}) with a critical boson solution, and we expect the quasi-Higgs phase to be replaced by a Higgs phase in which the $X$ boson condenses {\it i.e.\/} $\Delta_b = 0$. With $X$ condensed, a low frequency analysis of the saddle point equations shows that $\Delta_f = 1/2$. Consequently, such a Higgs phase realizes the disordered Fermi liquid, with electron correlations decaying as $1/\tau$, and spin correlations decaying as $1/\tau^2$.

\section{Renormalization group analysis}
\label{sec:RG}

This section returns to the problem as defined in Section~\ref{sec:largeN} for $M=2$. We will view the path integral in (\ref{Z}) as a quantum impurity problem in the presence of a bosonic bath $Q(\tau)$ and a fermionic bath $R(\tau)$; we defer imposition of the self-consistency conditions in (\ref{selfcons}). We will then follow the RG approach of Refs.~\onlinecite{FritzVojta04,FritzThesis} who studied a symmetric Anderson impurity coupled to a fermionic bath. The symmetric case is of relevance to us because we are considering the particle-hole symmetric case at half-filling, $p=0$. Our problem also has a bosonic bath, not present in the earlier work \cite{FritzVojta04,FritzThesis}, and we will include this bath using the methods of Ref.~\onlinecite{VBS2000}.

As we are looking for critical states, we assume that the fields $Q(\tau)$ and $R(\tau)$ have a power-law decay in time with
\begin{equation}
Q(\tau) \sim \frac{1}{|\tau|^{d-1}} \quad, \quad R(\tau) \sim \frac{\mbox{sgn}(\tau)}{|\tau|^{r+1}}\,. 
\label{QRpower}
\end{equation}
where, for now, $d$ and $r$ are arbitrary numbers determining exponents. Ultimately, once we have found a RG fixed point, the values of $d$ and $r$ can be fixed by using the self-consistency condition in (\ref{selfcons}). For now, our analysis exploits the freedom to choose $d$ and $r$: we will show that a systematic RG analysis of the path integral $\mathcal{Z}$ in (\ref{Z}) is possible in an expansion in $\epsilon$ and $\epsilon'$, where
\beq
\epsilon = 1-2r \quad, \quad \epsilon' = 2-d\,\,\,;\,\,\,
Q(\tau) \sim \frac{1}{|\tau|^{1-\epsilon'}} \quad, \quad R(\tau) \sim \frac{\mbox{sgn}(\tau)}{|\tau|^{(3-\epsilon)/2}}\,. 
\label{QRpower2}
\eeq
The analysis assumes $\epsilon$ and $\epsilon'$ are of the same order, and expands order-by-order in homogeneous polynomials in $\epsilon$ and $\epsilon'$.

We note that the perturbation expansion in power of $\epsilon$ is closely related to a weak-coupling small $U$ expansion of the symmetric Anderson model \cite{FritzVojta04}.
Consequently, the analysis is carried out directly in terms of the physical electron operator $c_\alpha$, and we will not fractionalize the electron into rotors and spinons, as we did in (\ref{defc}) for the large $M$ expansion in Section~\ref{sec:largeM}. It therefore possible that a critical point found in this approach is not `deconfined'. 

We proceed by decoupling the last two terms in the action $\mathcal{S}$ in (\ref{Z})  by introducing 
fermionic ($\psi_\alpha$) and bosonic ($\phi_a$, $a = x,y,z$) fields respectively, and then the 
path integral  reduces to a quantum impurity problem. The `impurity' is a single site of a particle-hole symmetric Hubbard model with 4 possible states, and this is coupled to the `bulk' $\psi_\alpha$ and $\phi_a$ excitations. The quantum impurity problem is specified by
the Hamiltonian
\begin{eqnarray}
H_{\rm imp} && = -\mu (n_{\uparrow} + n_{\downarrow}) + U n_{\uparrow} n_{ \downarrow} + g_0 \left( c_\alpha^\dagger \, \psi_\alpha (0) + \mbox{H.c.} \right) + \gamma_0 c_\alpha^\dagger \frac{\sigma^a_{\alpha\beta}}{2} c_\beta \, \phi_a (0) \nonumber \\
&&~~~~~~~ + \int |k|^r dk \, k \, \psi_{k\alpha}^\dagger \psi_{k \alpha} + \frac{1}{2} \int d^d x \left[ \pi_a^2 + (\partial_x \phi_a)^2 \right] \,. \label{Himp}
\end{eqnarray}
We now note features of the baths coupled to the impurity site. 

The bosonic bath is realized by a free massless scalar field in $d$ spatial dimensions,
as in Refs.~\cite{SBV1999,VBS2000,SS2001}. The field $\pi_a$ is canonically conjugate to the field $\phi_a$. 
The impurity spin $\vec{S}$ couples to the value of $\phi_a$ at the spatial origin, $\phi_a (0) \equiv 
\phi_a (x=0, \tau)$. It is easy to verify that upon integrating out $\phi_a$ from $H_{\rm imp}$, we obtain the $J$ term in $\mathcal{S}$, with $Q(\tau)$ obeying (\ref{QRpower}).

The fermionic bath is realized by free fermions $\psi_{k \alpha}$ with energy $k$ and a `pseudogap' 
density of states $\sim |k|^r$. The impurity electron operator $c_\alpha$ is coupled to $\psi_\alpha (0)  \equiv \int |k|^r dk \, \psi_{k \alpha}$. Integrating out $\psi_{k \alpha}$ from $H_{\rm imp}$ yields the $t$ term in $\mathcal{S}$, with $R(\tau)$ obeying (\ref{QRpower}).
 
It turns out that the structure of the RG shows that there is additional `boundary' renormalization of $\phi_a^2$ that must be accounted for in the $\epsilon, \epsilon'$ expansion, and this introduces a new coupling $\zeta$. This is an important distinction from the previous analysis of the $t$-$J$ model at non-zero $p$ \cite{Joshi2019}, where there were no additional boundary renormalizations; this boundary renormalization prevents us from making an all-orders extrapolation of certain exponents that was possible earlier \cite{Joshi2019}.
Our RG analysis will be carried out for the following action which includes the $\zeta$ coupling
\begin{align}
\mathcal{S}_c  =& \int \frac{d\omega}{2\pi} c_{\alpha}^{\dag}(\omega)(iA_{0}\textrm{sgn}(\omega)|\omega|^{r}) c_{\alpha}(\omega)+\frac{1}{2}\int d^{d}x d \tau [(\partial_{\tau}\phi_{a})^{2}+(\partial_{x}\phi)^{2}] \notag\\
&+\frac{U_{0}}{2}\int  d\tau 
c_{\alpha}^{\dag}c_{\beta}^{\dag}c_{\beta} c_{\alpha}+ \gamma_{0}\int d\tau c_{\alpha}^{\dag} \frac{\sigma^{a}_{\alpha\beta}}{2} c_{\beta} \phi_{a}(0)  +\frac{\zeta_{0}}{2} \int  d\tau \phi_{a}(0)^{2}\,, \label{act0}
\end{align}  
Note that we have already integrated out the fermionic bath $\psi_{k \alpha}$ (as was also done in Refs.~\onlinecite{FritzVojta04,FritzThesis}), to obtain a non-local propagator for the electron $c_\alpha$;
$A_0$ is an unimportant normalization constant which has absorbed the value of $g_0$. The action $\mathcal{S}_c$ contains three coupling constants, $U_0$, $\gamma_0$, and $\zeta_0$, and we will present their renormalization group flow below. 

\subsection{RG equations and fixed points}
\label{sec:RGe}

The RG equations of the action (\ref{act0}) are derived in Appendix~\ref{app:RG} for a generalized model with $\textrm{SU}(M)\times \textrm{SU}(M')$ spin symmetry. For the SU($M=2$) case and $M'=1$, and to the leading order, the flow equations from (\ref{rgoneloopM}) are (notice that in this case we define $\beta_{U} = (\beta_{U}+\beta_{V})|_{U+V\to U}$)
     \begin{align}
&\beta_{\gamma} = \frac{1}{2}(\epsilon-\epsilon')\gamma -\frac{\gamma U}{\pi A_{0}^{2}}+\frac{\zeta \gamma}{2\pi}\,,  \notag\\
& \beta_{U} = \epsilon U - \frac{3\gamma^{2}}{8\pi }\,, \notag\\
&\beta_{\zeta} = -\epsilon'\zeta +\frac{\zeta^{2}}{2\pi} +\frac{\gamma^{2}}{2\pi A_{0}^{2}} \,, \label{su2betas}
 \end{align}
where $\epsilon=1-2r$ and $\epsilon'=2-d$. Using the beta functions (\ref{su2betas}) we find the following four fixed points $(\gamma^{*2}, U^{*}, \zeta^{*})$:
    \begin{align}
    \label{fp1}
FP_{1} &= (0,0,0) \\
\label{fp2}
FP_{2} &= (0,0, 2\pi \epsilon') \\
\label{fp3}
FP_{3} &= \big(\frac{4\pi^{2}A_{0}^{2}}{9}\epsilon(\epsilon + \xi), \frac{\pi A_{0}^{2}}{6}(\epsilon + \xi),  \frac{\pi}{3}(3\epsilon'-2\epsilon + \xi) \big) \\
 \label{fp4}
FP_{4} &= \big(
\frac{4\pi^{2}A_{0}^{2}}{9}\epsilon(\epsilon - \xi),  
\frac{\pi A_{0}^{2}}{6}(\epsilon - \xi),  \frac{\pi}{3}(3\epsilon'-2\epsilon - \xi) \big)
 \end{align}
where $\xi =\sqrt{9\epsilon'^2-8\epsilon^{2}}$.
In order for the fixed points (\ref{fp3}) and (\ref{fp4}) to be real $\xi$ has to be real, which gives the condition $\epsilon'^2 >8\epsilon^2/9$. Additionally, for the fixed point (\ref{fp4}) to be real we should also have $\epsilon>\xi$, which gives $\epsilon'^2 <\epsilon^2$. 

We now analyze the stability of the fixed points by looking at the eigenvalues of the stability matrix (see Appendix \ref{app:RG} for details) and thus the RG flow. We will be interested in the situation when the non-trivial fixed points (\ref {fp3}) and (\ref{fp4}) are real, i.e., $8/9 < (\epsilon'/\epsilon)^{2} < 1$. We will have $\epsilon>0$, and discuss the situations when $\epsilon'$ is positive or negative.

\begin{figure}[t]
    \centering
    \subfloat[]{\includegraphics[width=0.3\textwidth]{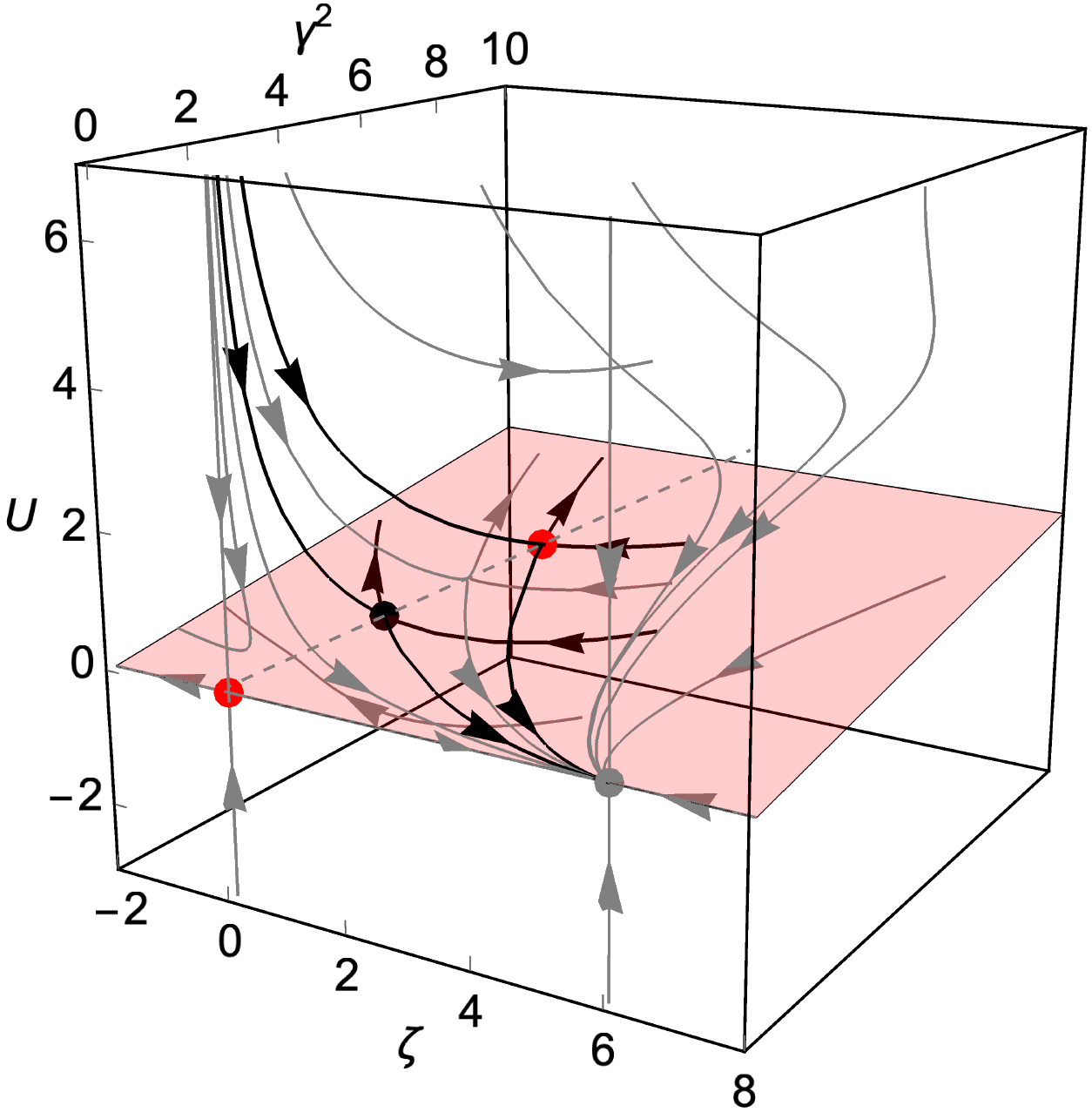}} ~~
    \subfloat[]{\includegraphics[width=0.3\textwidth]{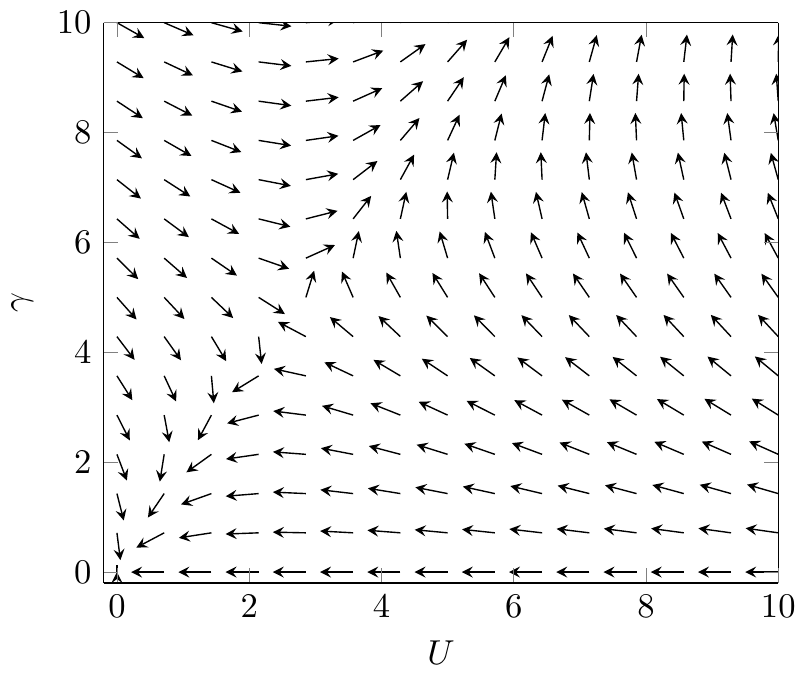}} ~~
    \subfloat[]{\includegraphics[width=0.3\textwidth]{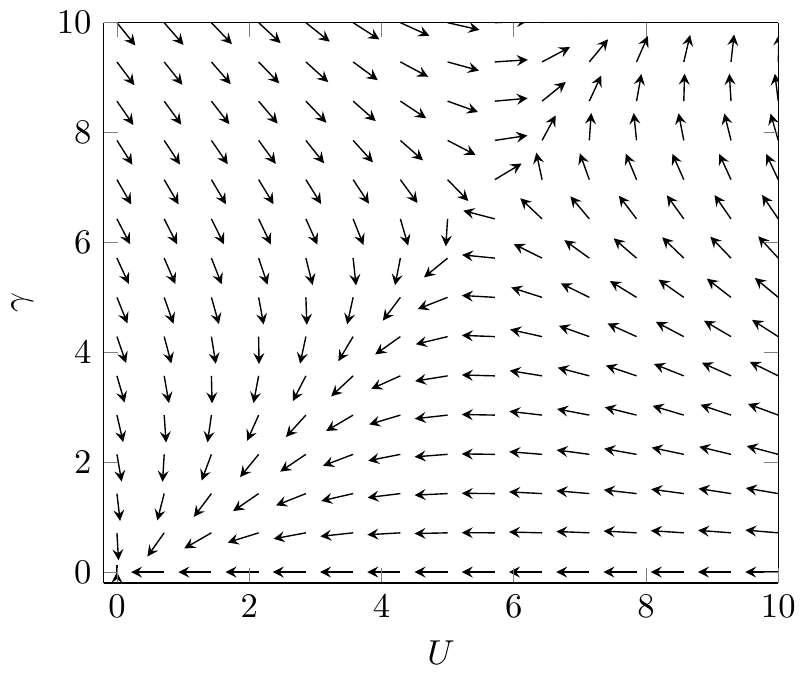}}
    \caption{(a) One-loop RG flow diagram in the $\zeta-\gamma^2-U$ space for $\epsilon=1$ and $\epsilon'=0.95$. The gray point is the stable fixed point ($FP_2$), red points are the fixed points ($FP_1$, $FP_3$) with one relevant direction, and the black point is the non-trivial fixed point ($FP_4$) with two relevant directions. All four fixed points are on the same plane $\frac{3\gamma^2}{8\pi\epsilon}-U=0$ coloured in light red. The black lines are the separatrix for the non-trivial fixed points. The dashed line connecting $FP_3$ and $FP_4$ has $\beta_\gamma=\beta_U=0$.
    (b) One-loop RG flow projected on to the $U-\gamma$ plane for $\epsilon'>0$. (c) Same as (b) for $\epsilon'<0$. In both (b) and (c) the nature of the flow is similar and controlled by the non-trivial fixed point $FP_3$ (\ref {fp3}), which corresponds to a quantum critical point.}
    \label{fig:rg_flow}
\end{figure}

(i) {\em $\epsilon'>0$}: In this case, the trivial fixed point (\ref {fp2}) is the only stable fixed point. The Gaussian fixed point (\ref {fp1}) and the non-trivial fixed point (\ref {fp3}) have one relevant direction and two irrelevant directions, while the other non-trivial fixed point (\ref {fp4}) has one irrelevant direction and two relevant directions in the RG flow phase-space of $(\gamma^{2}, U, \zeta)$. This is shown in Fig. \ref{fig:rg_flow} (a), with a 2d projection on the $U-\gamma$ plane shown in Fig. \ref {fig:rg_flow} (b). Therefore, the non-trivial fixed point (\ref {fp3}) separates the RG flow between large $\gamma$ or $U$ and small $\gamma$ or $U$, and thus corresponds to a quantum critical point. 
We discuss the anomalous dimensions of the spin and electron correlators at this fixed point in Sec. \ref {sec:scaling}.

(i) {\em $\epsilon'<0$}: In this case, the Gaussian fixed point now becomes the stable fixed point, while the other trivial fixed point (\ref {fp2}) now has one relevant and two irrelevant directions. The discussion of the non-trivial fixed points is the same as in case (i). The corresponding RG flow projected in the $U-\gamma$ plane is shown in Fig. \ref {fig:rg_flow} (c). Here again the fixed point (\ref {fp3}) separates the flow at large $\gamma$ or $U$ and small $\gamma$ or $U$, and therefore corresponds to a quantum critical point. 


\subsection{Scaling dimensions}
\label{sec:scaling}

A significant feature of the action (\ref{act0}) is that the quadratic term in the $c_\alpha$ does not get renormalized because of its non-analytic dependence on frequency. Consequently, the electron correlator has the same long-time decay as in the Gaussian theory
\beq
G_c (\tau) \sim \frac{\mbox{sgn}(\tau)}{|\tau|^{1-r}}
\eeq
and the scaling dimension of the electron operator is given exactly by
\beq
\mbox{dim}[c_\alpha] = (1-r)/2 = (1+\epsilon)/4\,. \label{dimc}
\eeq
This result is the same as that in the $t$-$J$ model analysis by Joshi {\it et al.\/} \cite{Joshi2019}. Comparing with (\ref{QRpower}) and the self-consistency condition in (\ref{selfcons}), we observe that the only possible self-consistent value is $r=0$ or $\epsilon=1$. Clearly, the $\epsilon$-expansion cannot be trusted at this large value in determining other features of the RG. Note that the electron scaling dimension in (\ref{dimc}) at $r=0$ agrees with the gapless boson solutions of the large $M$ analysis, where the scaling dimension of $c_\alpha$ is specified by (\ref{Gc}) and (\ref{Deltafb}).

Turning to the spin operator ${\bm S}$, we now find important differences from the $t$-$J$ model analysis \cite{Joshi2019}. The scaling dimension of ${\bm S}$ is not a RG invariant because of the boundary renormalization associated with the coupling $\zeta$. 
As explained in the  Appendix~\ref{app:RG} the spin operator ${\bm S}$ mixes with the boundary value of the bosonic bath field $\phi_{a}(0)$. This mixing is described by 
the matrix of anomalous dimensions is 
    \begin{align}
\gamma_{ij} = \left(  \begin{array}{cc} \displaystyle
  -\frac{U}{\pi A_{0}^{2}}   &  \displaystyle \frac{ \gamma}{2 \pi A_{0}^{2}} \\ 
   \displaystyle \frac{\gamma}{2\pi} & \displaystyle \frac{\zeta}{2\pi} \\ 
  \end{array}\right)\,.
 \end{align}
The full scaling dimensions are obtained by adding $\textrm{diag}(1-r, \frac{d-1}{2})$  to $\gamma_{ij}$
and diagonalizing the full matrix. At the fixed points we find,
\begin{align}
\label{del1}
FP_{1} &: \Delta_{+} =\frac{1 - \epsilon'}{2} \,, ~~
\Delta_{-} =\frac{1 + \epsilon}{2} \,, \\
\label{del2}
FP_{2} &: \Delta_{+} =\frac{1 + \epsilon'}{2} \,, ~~
\Delta_{-} =\frac{1 + \epsilon}{2} \,, \\
\label{del3}
FP_{3} &: \Delta_{+} =\frac{1+ \epsilon'}{2} \,, ~~
\Delta_{-} =\frac{1- \epsilon'}{2} \,, \\
\label{del4}
FP_{4} &: \Delta_{+} =\frac{1+ \epsilon'}{2} \,, ~~
\Delta_{-} =\frac{1- \epsilon'}{2} \,.
\end{align}
Note that these expressions are valid for both positive and negative values of $\epsilon'$, as long as the conditions mentioned above are met.

Let us use these scaling dimensions to impose the second of self-consistency conditions in (\ref{selfcons}). We consider the case of primary interest, the fixed point $FP_3$, with non-zero coupling. We have to match the exponent in (\ref{QRpower2}) with the smaller of the two exponents at $FP_3$; this leads us to conclude that self-consistency is achieved for any $\epsilon' \geq 0$. Presumably a specific value of $\epsilon'$ will be chosen at higher orders.

Finally, let us combine the consequences of the two self-consistency conditions in (\ref{selfcons}). From the first condition we found above that $\epsilon=1$. Using $\epsilon' \geq 0 $, 
we also have the restriction 
on the existence of $FP_3$ at real couplings, $\epsilon' > \sqrt{8/9} \epsilon$, and so the values of $\epsilon'$ are restricted to $\epsilon' > \sqrt{8/9} = 0.943$. We expect spin correlations to decay with time, and so (\ref{QRpower2}) also restricts $\epsilon' < 1$. But let us note that all these results are obtained as leading terms an expansion in $\epsilon$ and $\epsilon'$, so these large values cannot be trusted.

\subsection{Physical interpretation}

\begin{figure}[tb]
\begin{center}
\includegraphics[height=3.5in]{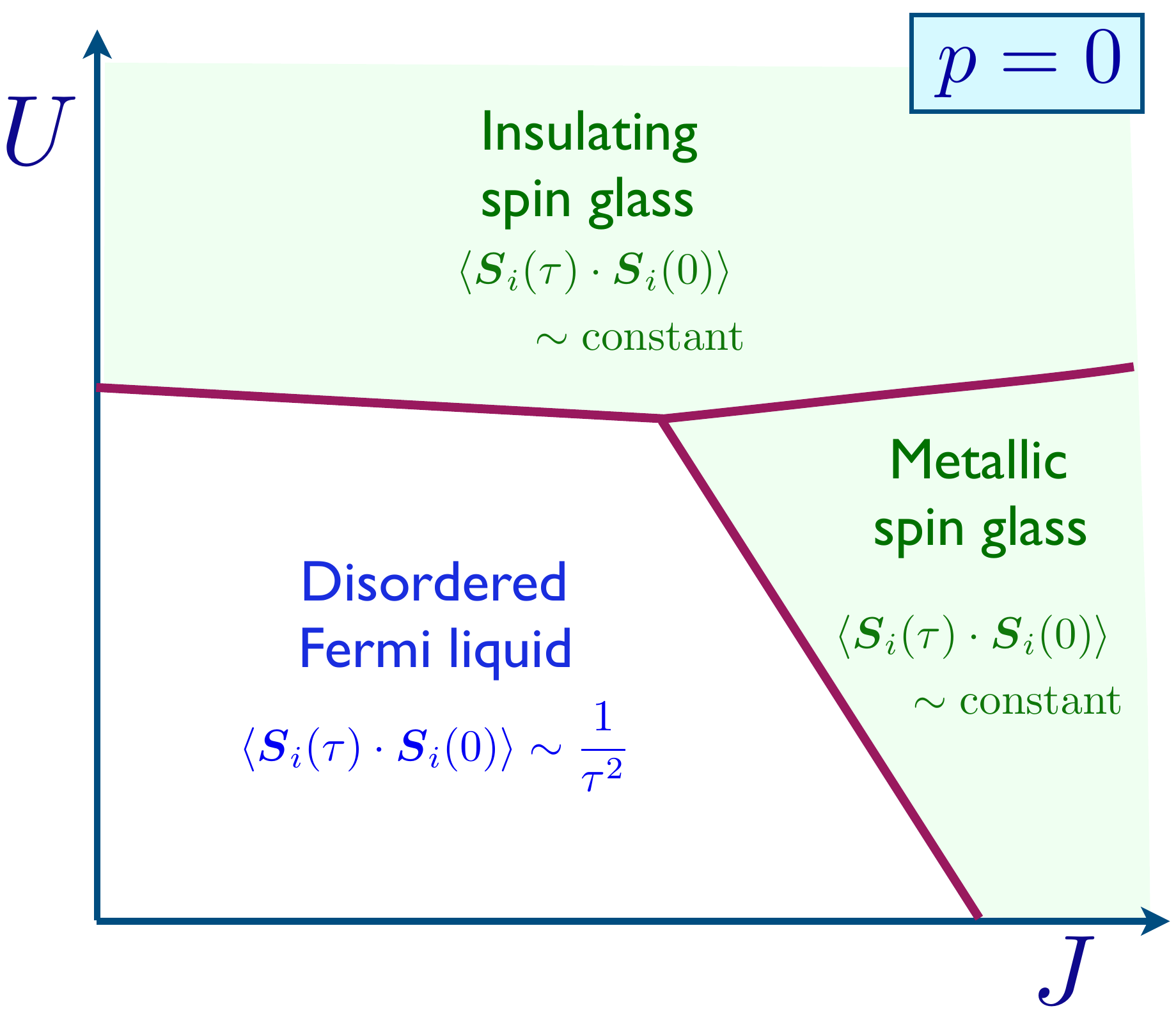} 
\end{center}
\caption{Schematic phase diagram at half-filling, $p=0$, as a function of the Hubbard repulsion, $U$, and the mean-square exchange interaction $J$. }
\label{fig:phasediag2}
\end{figure}

The physical interpretation of the quantum criticality described by the fixed point $FP_3$ remains an interesting open question. It is possible that it describes the deconfined metal-insulator transition at $p=0$ in Fig.~\ref{fig:phasediag}. However, another reasonable possibility is that the larger $U$ side of the fixed point is not an insulator, but a metallic spin glass. We sketch a possible phase diagram in Fig.~\ref{fig:phasediag2}, which shows a metallic spin glass phase at larger $J$ and small $U$.
One indication we are approaching a metallic spin glass phase is that the critical spin correlations at $FP_3$ decay extremely slowly. By (\ref{QRpower2}), the spin correlations decay with the exponent $(1-\epsilon')$, and we estimated above that $0< (1-\epsilon') < 0.057$ by (the unreliable) extrapolation from the one-loop results. 

In Appendix~\ref{app:metglass}, we will review an alternative Landau-type theory \cite{SRO1995,Sengupta95} of the transition to metallic spin glass order from a metallic phase. It is unlikely that the fixed point obtained in this section is the same as that of Appendix~\ref{app:metglass}: whereas the present RG analysis, when combined with the self-consistency conditions, requires $\epsilon' > 0$ on quite general grounds, we will see that the critical exponents in Appendix~\ref{app:metglass} require $\epsilon' < 0$.

\section{Conclusions}

Our paper has presented two approaches to analyzing the metal-insulator transition of the random Hubbard model (\ref{HU}) at half-filling ($p=0$); see Fig.~\ref{fig:phasediag}. 

The large $M$ analysis in Section~\ref{sec:largeM}
leads to the fractionalization of the electron $c_\alpha$ into a boson $X$ (with scaling dimension $\Delta_b$) and a fermion $f$ (with scaling dimension $\Delta_f$). We argued that the deconfined critical point with $\Delta_f = \Delta_b = 1/4$ realized the metal-insulator transition observed numerically by Cha {\it et al.} \cite{Cha19}.
These exponents imply the electron correlator in (\ref{Gc}) and the spin correlator in (\ref{Sc}), both consistent with the results of Cha {\it et al.} \cite{Cha19}.

We presented a renormalization group analysis of the random Hubbard model in Section~\ref{sec:RG}. 
This turns out to be effectively a small $U$ analysis, and to leading order in an $\epsilon$ expansion we found a fixed point ($FP_3$) possibly describing a metal-insulator transition. The renormalization group results have to be supplemented by the imposition of a self-consistency relation on the exponents, and we found that this required 
extrapolation to values of $\epsilon$ of order unity where our expansion breaks down. Consequently, we are not able to obtain the exponents accurately in this approach. We also discussed the possibility that $FP_3$ described the onset of a metallic spin glass phase from a disordered Fermi liquid, and that $FP_3$ was an alternative to the Landau theory of such a transition \cite{SRO1995,Sengupta95}.

\subsection*{Acknowledgements}

We thank P.~Cha, A.~Georges, O.~Parcollet, and M.~Vojta  
for valuable discussions.
This research was supported by the National Science Foundation under Grant No. DMR-1664842, by DOE grant DE-SC0019030, and by the CIFAR Quantum Materials program. 
G.T.  acknowledges support from the MURI grant W911NF-14-1-0003 from ARO and by DOE grant DE-SC0007870.

\appendix
\section{Derivation of RG}
\label{app:RG}
We will consider here a generalized model in which the spin index $\alpha = 1, \ldots ,M$, and the electron operator $c_{p,\alpha}$ has an additional `orbital' index $p = 1, \ldots, M'$. The latter index will be useful in the large $M$ analysis presented in Section~\ref{sec:largeM}. The  effective action generalizing (\ref{act0}) is
\begin{align}
\mathcal{S}_c  =& \int \frac{d\omega}{2\pi} c_{p,\alpha}^{\dag}(\omega)(iA_{0}\textrm{sgn}(\omega)|\omega|^{r}) c_{p,\alpha}(\omega)+\frac{1}{2}\int d^{d}x d \tau [(\partial_{\tau}\phi_{a})^{2}+(\partial_{x}\phi)^{2}]+ \gamma_{0}\int d\tau c_{p,\alpha}^{\dag} T^{a}_{\alpha\beta}c_{p,\beta} \phi_{a}(0) \notag\\
&+\frac{U_{0}}{2}\int  d\tau 
c_{p,\alpha}^{\dag}c_{q,\beta}^{\dag}c_{q,\beta} c_{p,\alpha}+\frac{V_{0}}{2}\int  d\tau 
c_{p,\alpha}^{\dag}c_{q,\beta}^{\dag}c_{p,\beta} c_{q,\alpha}  +\frac{\zeta_{0}}{2} \int  d\tau \phi_{a}(0)^{2}\,, \label{act1}
\end{align}  
where $p=1,\dots, M'$, $a=1, \dots, M^{2}-1$ and $\alpha =1,\dots, M$. We have introduced a coupling $V$, an additional local interaction allowed by the enlarged symmetry. The matrix $T^{a}$ is the fundamental representation of SU($M$) algebra, and has the properties
\begin{align}
\textrm{tr}(T^{a}T^{b}) = \frac{1}{2}\delta^{ab}, \quad T^{a}T^{a} = \frac{M^{2}-1}{2M} \cdot \textbf{1}\,, \quad T^{a}_{\alpha\beta}T^{a}_{\gamma\delta}=\frac{1}{2}
\Big(\delta_{\alpha\delta}\delta_{\beta \gamma}-\frac{1}{M}\delta_{\alpha\beta}\delta_{\gamma\delta}\Big)\,.
\end{align} 
The action (\ref{act1}) has $\textrm{SU}(M)\times \textrm{SU}(M')$ global symmetry. 

\begin{figure}[h!]
\begin{center}
\includegraphics[height=1in]{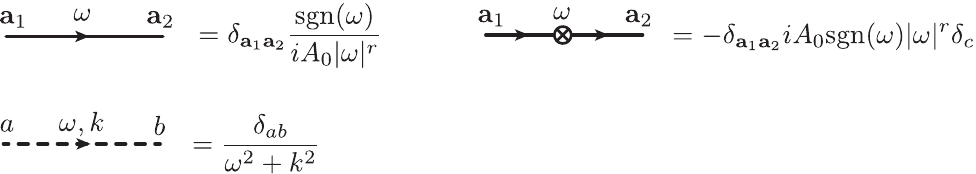} 
\end{center}
\caption{Feynman rules for the propagators} 
\label{fig:feynmprop}
\end{figure}

\begin{figure}[h!]
\begin{center}
\includegraphics[height=2.3in]{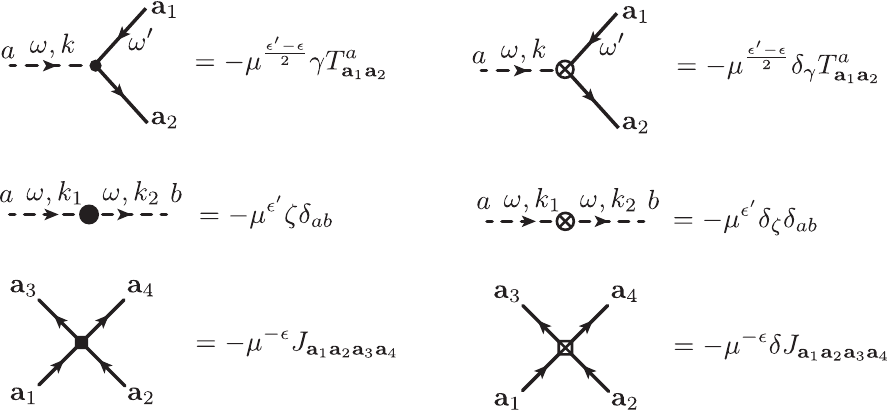} 
\end{center}
\caption{Feynman rules for the vertices} 
\label{fig:feynmvert}
\end{figure}

Dimensional analysis yields
\begin{align}
 [c]=[\mu]^{\frac{1-r}{2}}, \quad  [U_{0}] =[V_{0}] = [\mu]^{2r-1}, \quad [\gamma_{0}] = [\mu]^{\frac{1}{2}+r -\frac{d}{2}}, \quad \quad [\zeta_{0}] = [\mu]^{2-d}\,.
\end{align} 
Therefore introducing new notations $\epsilon=1-2r$ and $\epsilon'=2-d$ we write renormalized fields as $c_{0} = Z_{c}^{1/2}c$ and  $\phi_{0}=Z_{\phi}^{1/2} \phi$ and dimensionless couplings as 
\begin{align}
 U_{0} = \mu^{-\epsilon} Z_{c}^{-2}(U+\delta_{U})\,, \quad V_{0} = \mu^{-\epsilon} Z_{c}^{-2}(V+\delta_{V}), \quad \gamma_{0}=\mu^{\frac{\epsilon'-\epsilon}{2}} Z_{c}^{-1}Z_{\phi}^{-1/2}(\gamma+\delta_{\gamma})\,, \quad \zeta_{0} = \mu^{\epsilon'}Z_{\phi}^{-1}(\zeta+\delta_{\zeta})\,,
\end{align} 
where $Z_{\phi}= 1+\delta_{\phi}$, $Z_{c}=1+\delta_{c}$ and $\mu$ is a mass scale parameter.
\begin{align}
S  =& \int \frac{d\omega}{2\pi} c_{\textbf{a}}^{\dag}(\omega)(iA_{0}\textrm{sgn}(\omega)|\omega|^{r}) c_{\textbf{a}}(\omega)+\frac{1}{2}\int d^{d}x d \tau [(\partial_{\tau}\phi_{a})^{2}+(\partial_{x}\phi)^{2}]+  \mu^{\frac{\epsilon'-\epsilon}{2}} \gamma \int d\tau c^{\dag}_{\textbf{a}_{1}}T^{a}_{\textbf{a}_{1}\textbf{a}_{2}}c_{\textbf{a}_{2}}\phi_{a}   \notag\\
&+\frac{1}{4} \mu^{-\epsilon} \int d\tau J_{\textbf{a}_{1}\textbf{a}_{2},\textbf{a}_{3}\textbf{a}_{4}}c_{\textbf{a}_{1}}^{\dag}c_{\textbf{a}_{2}}^{\dag}c_{\textbf{a}_{3}}c_{\textbf{a}_{4}} +\frac{1}{2}\mu^{\epsilon'}\zeta \int  d\tau \phi_{a}(0,\tau)^{2}+\delta S\,, \label{renormS}
\end{align}  
where the counter-terms action reads 
\begin{align}
\delta S  =&  \delta_{c}\int \frac{d\omega}{2\pi} c_{\textbf{a}}^{\dag}(\omega)(iA_{0}\textrm{sgn}(\omega)|\omega|^{r}) c_{\textbf{a}}(\omega)
+\frac{\delta_{\phi}}{2}\int d^{d}x d \tau [(\partial_{\tau}\phi_{a})^{2}+(\partial_{x}\phi)^{2}] +  \mu^{\frac{\epsilon'-\epsilon}{2}} \delta_{\gamma} \int d\tau c^{\dag}_{\textbf{a}_{1}}T^{a}_{\textbf{a}_{1}\textbf{a}_{2}}c_{\textbf{a}_{2}}\phi_{a} 
\notag\\
&+\frac{1}{4} \mu^{-\epsilon} \int d\tau \delta J_{\textbf{a}_{1}\textbf{a}_{2},\textbf{a}_{3}\textbf{a}_{4}}c_{\textbf{a}_{1}}^{\dag}c_{\textbf{a}_{2}}^{\dag}c_{\textbf{a}_{3}}c_{\textbf{a}_{4}} +\frac{1}{2}\mu^{\epsilon'}\delta_{\zeta} \int  d\tau \phi_{a}(0,\tau)^{2} 
\end{align}  
and we combined two indices $(p,\alpha)$ in one bold index $\textbf{a}$  and introduce  $T^{a}_{\textbf{a}_{1}\textbf{a}_{2}} = \delta_{p_{1}p_{2}}T^{a}_{\alpha_{1}\alpha_{2}}$ and the tensor $J_{\textbf{a}_{1}\textbf{a}_{2},\textbf{a}_{3}\textbf{a}_{4}}$, which reads  $J_{\textbf{a}_{1}\textbf{a}_{2},\textbf{a}_{3}\textbf{a}_{4}} = U J^{(1)}_{\textbf{a}_{1}\textbf{a}_{2},\textbf{a}_{3}\textbf{a}_{4}}+
V J^{(2)}_{\textbf{a}_{1}\textbf{a}_{2},\textbf{a}_{3}\textbf{a}_{4}}$ and is given by
\begin{align}
J_{\textbf{a}_{1}\textbf{a}_{2},\textbf{a}_{3}\textbf{a}_{4}} =U(\delta_{p_{1}p_{4}}\delta_{\alpha_{1}\alpha_{4}} \delta_{p_{2}p_{3}}\delta_{\alpha_{2}\alpha_{3}}-
\delta_{p_{1}p_{3}}\delta_{\alpha_{1}\alpha_{3}} \delta_{p_{2}p_{4}}\delta_{\alpha_{2}\alpha_{4}})+
V(\delta_{p_{1}p_{3}}\delta_{\alpha_{1}\alpha_{4}} \delta_{p_{2}p_{4}}\delta_{\alpha_{2}\alpha_{3}}-
\delta_{p_{1}p_{4}}\delta_{\alpha_{1}\alpha_{3}} \delta_{p_{2}p_{3}}\delta_{\alpha_{2}\alpha_{4}})\,.
\end{align} 
We depicted Feynman rules for the propagators in figure \ref{fig:feynmprop} and for the vertices in figure \ref{fig:feynmvert}.
Next we are computing all diagrams contributing in the leading order to renormalization 
of couplings $\gamma$, $U$, $V$ and $\zeta$.

\begin{figure}[h!]
\begin{center}
\includegraphics[height=0.8in]{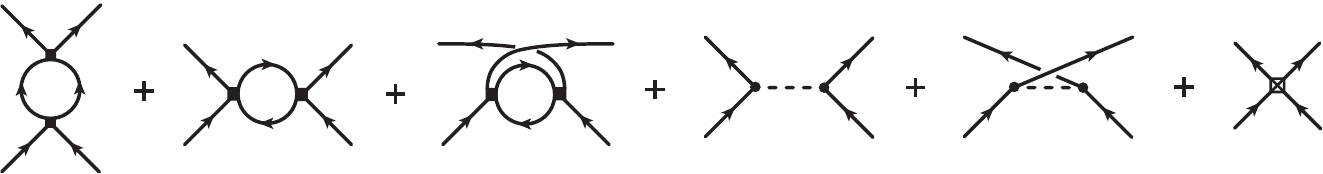} 
\end{center}
\caption{Leading order diagrams contributing to $\beta_{U}$ and $\beta_{V}$.} 
\label{fig:oneloopJ}
\end{figure}

\begin{figure}[h!]
\begin{center}
\includegraphics[height=0.55in]{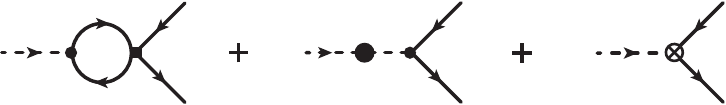} 
\end{center}
\caption{Leading order diagrams contributing to $\beta_{\gamma}$. }
\label{fig:oneloopG}
\end{figure}

\begin{figure}[h!]
\begin{center}
\includegraphics[height=0.4in]{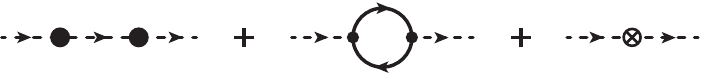} 
\end{center}
\caption{Leading order diagrams contributing to $\beta_{\zeta}$. }
\label{fig:oneloopzeta}
\end{figure}

\begin{figure}[h!]
\begin{center}
\includegraphics[height=0.4in]{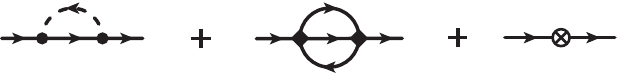} 
\end{center}
\caption{Leading order diagrams contributing to renormalization of the fermionic self energy.} 
\label{fig:oneloopself}
\end{figure}

\begin{figure}[h!]
\begin{center}
\includegraphics[height=0.6in]{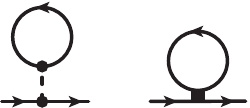} 
\end{center}
\caption{An example of bubble Feynman diagrams, which are  zero and omitted in all other pictures.} 
\label{fig:bubbles}
\end{figure}

 We assume that bubble diagrams like in figure \ref{fig:bubbles} are zero and omit them in all the figures. 
To compute the Feynman diagrams we use three main integrals  
\begin{align}
&\int_{-\infty}^{+\infty} \frac{d \omega}{2\pi}\frac{\textrm{sgn}(\omega)\textrm{sgn}(\omega+\omega')}{|\omega|^{\alpha_{1}}|\omega+\omega'|^{\alpha_{2}}} = \frac{1}{(4\pi)^{1/2}}\frac{\Gamma (\frac{2-\alpha_{1}}{2}) \Gamma (\frac{2-\alpha_{2}}{2}) \Gamma (\frac{\alpha_{1}+\alpha_{2}-1}{2} )}{\Gamma (\frac{1+\alpha_{1}}{2}) \Gamma (\frac{1+\alpha_{2}}{2}) \Gamma (\frac{2-\alpha_{1}-\alpha_{2}}{2} )}  \frac{1}{|\omega'|^{\alpha_{1}+\alpha_{2}-1}}\,,\notag\\
&\int_{-\infty}^{+\infty} \frac{d \omega}{2\pi}\frac{\textrm{sgn}(\omega+\omega')}{|\omega+\omega'|^{\alpha_{1}}|\omega|^{\alpha_{2}}} =\frac{1}{(4\pi)^{1/2}}\frac{\Gamma (\frac{2-\alpha_{1}}{2}) \Gamma (\frac{1-\alpha_{2}}{2}) \Gamma (\frac{\alpha_{1}+\alpha_{2}}{2} )}{\Gamma (\frac{1+\alpha_{1}}{2}) \Gamma (\frac{\alpha_{2}}{2}) \Gamma (\frac{3-\alpha_{1}-\alpha_{2}}{2} )}\frac{\textrm{sgn}(\omega')}{|\omega'|^{\alpha_{1}+\alpha_{2}-1}} \,, \notag\\
&\int \frac{d^{d}k}{(2\pi)^{d}}\frac{k^{\alpha}}{k^{2}+\omega^{2}} =\frac{\pi}{(4\pi)^{d/2} \Gamma(d/2) \sin \frac{\pi(d+\alpha)}{2}}\frac{1}{|\omega|^{2-d-\alpha}}\,. \label{mainintegrals}
\end{align} 
 Using these integrals we find for the  leading order counter-terms 
   \begin{align}
&\delta_{\gamma}^{(1)} = \frac{\gamma (U+M' V)}{\pi A_{0}^{2}  }\frac{1}{\epsilon} + \frac{\zeta \gamma}{2\pi}\frac{1}{\epsilon'} \,, \notag\\
& \delta_{U}^{(1)} = -\frac{U^{2}(MM'-2)+2(M-M')U V -V^{2}}{\pi A_{0}^{2}}\frac{1}{\epsilon} - \frac{\gamma^{2}}{4\pi M}\frac{1}{\epsilon'}\,,  \notag\\
&\delta_{V}^{(1)} =-\frac{V^{2}(M-M')-2U V}{ \pi A_{0}^{2} }\frac{1}{\epsilon}  - \frac{\gamma^{2}}{4\pi }\frac{1}{\epsilon'} \,, \notag\\
& \delta_{\zeta}^{(1)} = \frac{\zeta^{2}}{2\pi}\frac{1}{\epsilon'} -\frac{M' \gamma^{2}}{2 \pi A_{0}^{2}}\frac{1}{\epsilon} \,.
 \end{align}
 and $\delta_{c}^{(1)}=0$ and also $\delta_{\phi}=0$ at all orders. 
Using that 
 $\beta_{\gamma}= d \gamma/ d\log \mu$,  $\beta_{U}= d U /d\log \mu  $, $\beta_{V}= d V /d\log \mu  $  and $\beta_{\zeta} =  d \zeta/ d\log \mu$ we find the leading order beta functions
     \begin{align}
&\beta_{\gamma} = \frac{1}{2}(\epsilon-\epsilon')\gamma -\frac{\gamma (U+M' V)}{\pi A_{0}^{2}}+\frac{\zeta \gamma}{2\pi}\,,  \notag\\
& \beta_{U} = \epsilon U +\frac{U^{2}(MM'-2)+2(M-M')U V -V^{2}}{\pi A_{0}^{2}} - \frac{\gamma^{2}}{4\pi M}\,, \notag\\
&\beta_{V} =\epsilon V +\frac{V^{2}(M-M')-2U V}{\pi A_{0}^{2}} - \frac{\gamma^{2}}{4\pi } \,,\notag\\
&\beta_{\zeta} =-\epsilon' \zeta +\frac{\zeta^{2}}{2\pi} +\frac{M'\gamma^{2}}{2\pi A_{0}^{2}}\,.
\label{rgoneloopM}
 \end{align}

\subsection{Anomalous dimension of the operator $S_{a}=c^{\dag}_{\textbf{a}_{1}}T^{a}_{\textbf{a}_{1}\textbf{a}_{2}}c_{\textbf{a}_{2}}$}
\begin{figure}[h!]
\includegraphics[width=0.6\textwidth]{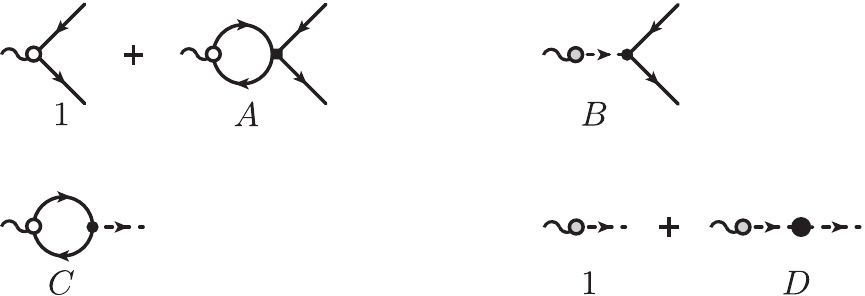}
\caption{\label{ABCD} All leading order diagrams for renormalization of operators $S_{a}$ and $\phi_{a}$.}
\end{figure}

 We would like to compute anomalous dimension of the operator  $S_{a}(\tau)=c^{\dag}_{\textbf{a}_{1}}(\tau)T^{a}_{\textbf{a}_{1}\textbf{a}_{2}}c_{\textbf{a}_{2}}(\tau)$. 
 First of all we notice that it has  bare dimension $[S_{a}]=[\mu]^{1-r}$. On the other hand the operator $\phi_{a}(\tau,x=0)$ has  bare dimension 
 $[\phi_{a}]=[\mu]^{\frac{d-1}{2}}$, therefore when $r\to 1/2$ and $d\to 2$ these two operators can mix. Thus we define renormalized dimensionless operators as
 \begin{align}
 &[S_{a}(\tau)]_{R} = Z_{SS} \mu^{r-1}S_{a}(\tau)+Z_{S\phi} \mu^{\frac{1-d}{2}}\phi_{a}(\tau,0)\,, \notag\\
 &[\phi_{a}(\tau)]_{R}= Z_{\phi S}  \mu^{r-1}S_{a}(\tau)+Z_{\phi \phi}\mu^{\frac{1-d}{2}}\phi_{a}(\tau,0)\,,
 \end{align}
 where the counter-terms have only poles in $\epsilon$ and $\epsilon'$ and have the following form
  \begin{align}
&Z_{SS} = 1 +z_{SS}^{(1)}(U,V,\gamma,\zeta,\epsilon,\epsilon')+z_{SS}^{(2)}(U,V,\gamma,\zeta,\epsilon,\epsilon')+\dots\,, \notag\\
&Z_{S\phi} = z_{S\phi}^{(1)}(U,V,\gamma,\zeta,\epsilon,\epsilon')+z_{S\phi}^{(2)}(U,V,\gamma,\zeta,\epsilon,\epsilon')+\dots\,, \notag\\
&Z_{\phi S} = z_{\phi S}^{(1)}(U,V,\gamma,\zeta,\epsilon,\epsilon')+z_{\phi S}^{(2)}(U,V,\gamma,\zeta,\epsilon,\epsilon')+\dots\,, \notag\\
&Z_{\phi\phi} = 1 +z_{\phi\phi}^{(1)}(U,V,\gamma,\zeta,\epsilon,\epsilon')+z_{\phi\phi}^{(2)}(U,V,\gamma,\zeta,\epsilon,\epsilon')+\dots\,,
 \end{align}
 and as usual $\epsilon=1-2r$ and $\epsilon'=2-d$. It is convenient to work in momentum space so we have 
  \begin{align}
 &[S_{a}(\omega)]_{R} = Z_{SS} \mu^{r-1} S_{a}(\omega)+Z_{S\phi} \mu^{\frac{1-d}{2}} \int \frac{d^{d}k}{(2\pi)^{d}}\phi_{a}(\omega,k)\,.
 \end{align}
 We find the counter-terms $Z_{SS}$, $Z_{S\phi}$, $Z_{\phi S}$ and $Z_{\phi \phi}$ by demanding that expressions 
   \begin{align}
& \Gamma_{SS}(\omega|\omega') = \frac{\langle [S_{a}(\omega)]_{R} c_{\textbf{a}_{1}}(\omega') c^{\dag}_{\textbf{a}_{2}}(\omega+\omega')\rangle_{c}  }{G(\omega')G(\omega+\omega')}, \quad 
  \Gamma_{S\phi}(\omega|k) = \frac{\langle [S_{a}(\omega)]_{R} \phi_{b}(-\omega,-k)\rangle_{c}  }{D(\omega,k)}\,,   \notag\\
 & \Gamma_{\phi S}(\omega|\omega') = \frac{\langle [\phi_{a}(\omega)]_{R} c_{\textbf{a}_{1}}(\omega') c^{\dag}_{\textbf{a}_{2}}(\omega+\omega')\rangle_{c}  }{G(\omega')G(\omega+\omega')}, \quad 
  \Gamma_{\phi\phi}(\omega|k) = \frac{\langle [\phi_{a}(\omega)]_{R} \phi_{b}(-\omega,-k)\rangle_{c}  }{D(\omega,k)}
  \label{3ptfunctions}
 \end{align}
 are free of divergencies, where $G(\omega)=\langle c(\omega)c^{\dag}(\omega)\rangle$ and $D(\omega,k)=\langle \phi(\omega,k)\phi(-\omega,-k)\rangle=1/(\omega^{2}+k^{2})$. To the leading order we can write expressions for  (\ref{3ptfunctions}) as 
   \begin{align}
 &\Gamma_{SS}(\omega|\omega') =Z_{SS}  \mu^{r-1}(1+A)+
 Z_{S\phi}\mu^{\frac{1-d}{2}}B\,,\notag\\
 &\Gamma_{S\phi}(\omega|k)= Z_{SS} \mu^{r-1} C+Z_{S\phi}\mu^{\frac{1-d}{2}}(1+D)\,, \notag\\
  &\Gamma_{\phi S}(\omega|\omega') =Z_{\phi S}  \mu^{r-1}(1+A)+
 Z_{\phi \phi}\mu^{\frac{1-d}{2}}B\,,\notag\\
 &\Gamma_{\phi \phi}(\omega|k)= Z_{\phi S} \mu^{r-1} C+Z_{\phi \phi}\mu^{\frac{1-d}{2}}(1+D)\,,
 \end{align}
where diagrams $A, B, C$ and $D$ are listed in figure \ref{ABCD}.
Computing the diagrams we find the counter-terms 
   \begin{align}
z_{ss}^{(1)}= \frac{U+M' V}{\pi A_{0}^{2}}\frac{1}{\epsilon}, \quad z_{s\phi}^{(1)}= -\frac{M' \gamma}{2 \pi A_{0}^{2}}\frac{1}{\epsilon} \quad z_{\phi s}^{(1)} = 
\frac{\gamma}{2\pi}\frac{1}{\epsilon'}, \quad z_{\phi \phi}^{(1)} =  \frac{\zeta}{2\pi }\frac{1}{\epsilon'}\,.
 \end{align}
Therefore the matrix of anomalous dimensions is
    \begin{align}
\gamma_{ij} = \left(  \begin{array}{cc}
  -\frac{U+M' V}{\pi A_{0}^{2}}   & \frac{M' \gamma}{2 \pi A_{0}^{2}} \\ 
   \frac{\gamma}{2\pi} & \frac{\zeta}{2\pi} \\ 
  \end{array}\right)\,.
 \end{align}
The full scaling dimensions are obtained by adding $\textrm{diag}(1-r, \frac{d-1}{2})$  to $\gamma_{ij}$
and diagonalizing the full matrix. 

\subsection{Stability of fixed points}
In this subsection we discuss the stability of fixed points found above and listed in Eqs. (\ref{fp1}) - (\ref{fp4}). First we construct the stability matrix,
\begin{equation}
\label{eq_J_mat}
J \equiv 
\begin{bmatrix}
J_{1} & J_{2} & J_{3} \\
J_{4} & J_{5} & J_{6} \\
J_{7} & J_{8} & J_{9} 
\end{bmatrix} \,,
\end{equation}
where,
\begin{align}
\label{eq:Jdef}
J_{1} &\equiv \frac{\partial \beta_\gamma}{\partial \gamma} = \frac{\epsilon-\epsilon'}{2}-\frac{U}{\pi A_{0}^{2}}+\frac{\zeta}{2\pi} \,, ~~~~
J_{2} \equiv \frac{\partial \beta_\gamma}{\partial U} = -\frac{\gamma}{\pi A_{0}^{2}} \,, ~~~~ 
J_{3} \equiv \frac{\partial \beta_\gamma}{\partial \zeta} = \frac{\gamma}{2\pi} \,, \nonumber \\ 
J_{4} &\equiv \frac{\partial \beta_U}{\partial \gamma} = -\frac{3\gamma}{4\pi} \,, ~~~~
J_{5} \equiv \frac{\partial \beta_U}{\partial U} = \epsilon \,, ~~~~
J_{6} \equiv \frac{\partial \beta_U}{\partial \zeta} = 0 \,, \nonumber \\
J_{7} &\equiv \frac{\partial \beta_\zeta}{\partial \gamma} = \frac{\gamma}{\pi A_{0}^{2}} \,, ~~~~
J_{8} \equiv \frac{\partial \beta_\zeta}{\partial U} = 0 \,, ~~~~
J_{9} \equiv \frac{\partial \beta_\zeta}{\partial \zeta} = -\epsilon' + \frac{\zeta}{\pi} \,.
\end{align}
The eigenvalues of matrix $J$ at the trivial fixed points (\ref{fp1}) and (\ref{fp2}) are $(\epsilon,(\epsilon-\epsilon')/2,-\epsilon')$ and $(\epsilon,(\epsilon+\epsilon')/2,\epsilon')$ respectively. Thus, for any positive values of $\epsilon$ and $\epsilon'$ all the eigenvalues corresponding to $FP_{2}$ are positive, and thus it is a stable fixed point. The eigenvalues at the non-trivial fixed points (\ref{fp3}) and (\ref{fp4}) are given by the roots of their respective characteristic polynomials, 
\begin{equation}
\label{p34}
p_{3}(\lambda) = \lambda^{3} + a_{3} \lambda^{2} + b_{3}\lambda + c_{3} \,, ~~~    
p_{4}(\lambda) = \lambda^{3} + a_{4} \lambda^{2} + b_{4}\lambda + c_{4} \,,
\end{equation}
where,
\begin{equation}
\label{abc}
a_{3,4}=-\frac{\epsilon \pm \xi}{3} \,, ~~~
b_{3,4}=\frac{\epsilon}{9}(-11\epsilon \mp 2\xi) \,, ~~~
c_{3,4}=\pm\frac{2}{9}\epsilon \xi (\epsilon \pm \xi) \,.
\end{equation}
Recall that for non-trivial fixed points to be real we need $\xi$ to be real as well as $\epsilon > \xi$. Now examining the coefficients of the characteristic polynomials we see that the fixed point (\ref{fp3}) has one negative eigenvalue and two positive eigenvalues, while (\ref{fp4}) has two negative eigenvalues and one positive eigenvalue. 

 \subsection{Next order beta functions}
\begin{figure}[h!]
\includegraphics[width=1.\textwidth]{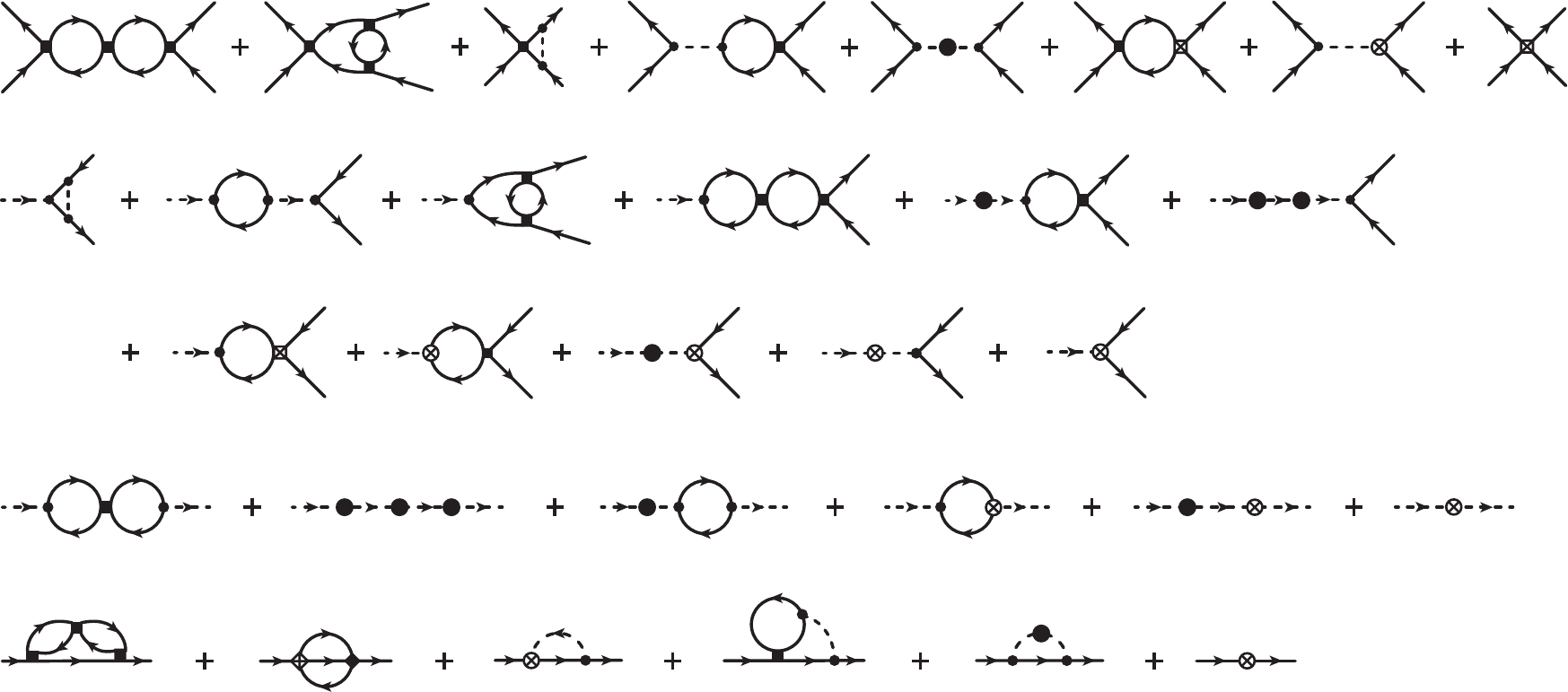}
\caption{ The next order diagrams contributing to renormalization of the couplings and   self-energy. }
\label{twoloopdiags}
\end{figure}

 In this subsection we list all diagrams in figure \ref{twoloopdiags} contributing at the next order to the beta functions and renormalization of the fermionic self-energy. Notice that we list only diagrams which represent distinct  graphs, whereas some diagrams in this list can include different sub-cases, like in figure \ref{fig:oneloopJ}. 
All the diagrams can be computed with the use of the integrals in eq. (\ref{mainintegrals}). The result of the computation is 
\begin{align}
 \beta_{U}^{(2)} =&-\frac{ \big( (M M'+1) U^3+(M M'+5) U V^2+(M-M') (2 U^2 V-V^3) \big)(\pi -4 \log (2))}{\pi ^2 A_{0}^4}\notag\\
 &-\frac{ (M^2-1)\gamma ^2 U (\gamma_{E} -\log (4 \pi ))}{4 \pi ^2 A_{0}^2 M} \,,\notag\\
 \beta_{V}^{(2)} =&-\frac{V \big((M M'+5) U^2 +2 M M' V^2+2  (M-M') U V\big)(\pi -4 \log (2)) }{\pi ^2 A_{0}^4}-\frac{ (M^2-1) \gamma ^2 V (\gamma_{E} -\log (4 \pi ))}{4 \pi ^2 A_{0}^2 M} \,, \notag\\
 \beta_{\gamma}^{(2)}=& -\frac{ \big((M M'-1) (U^2+V^2)+2(M-M')U V\big)\gamma  (\pi -4 \log (2))}{2 \pi^2 A_{0}^4}
 +\frac{\gamma^3 (\gamma_{E}-\log (4 \pi ) )}{8 \pi^2 A_{0}^{2} M}\,,
 \end{align}
 and $\beta_{\zeta}^{(2)} =0$ and $\delta_{c}^{(2)}=0$, where $\gamma_{E}$ is the Euler constant. We see that there is no renormalization of the fermionic self-energy to this order and we can guess that this is true to all orders of the perturbation theory.  We also notice that upon taking $\gamma=0$, $M=2$, $M'=1$ and defining $\beta_{U} = (\beta_{U}+\beta_{V})|_{U+V\to U}$ we recover the result reported in  \cite{FritzVojta04}.

 \section{Onset of spin glass order in a metal}
\label{app:metglass}

This Appendix will address the transition from the disordered Fermi liquid phase to the metallic spin glass. There are 3 possible theories of this transition:\\
({\it i}) At large $U$, and for $p>0$, eliminate the doubly-occupied site, and address the transition in the $t$-$J$ model. This yields a deconfined critical point, described in Ref.~\onlinecite{Joshi2019}.\\
({\it ii\/}) At small $U$, and at $p=0$, perform the RG analysis presented in Section~\ref{sec:RG}. This yields fixed point $FP_3$, which could describe the onset of spin glass order in a metal.\\
({\it iii\/}) Use a weak-coupling Landau functional approach to quantum spin glasses \cite{RSY95,SRO1995}, which we review in this section.

We begin with the disordered-averaged imaginary time path integral of the Hamiltonian in (\ref{HU}), keeping track of replica indices, $a,b=1 \ldots n$; at the end we need to take the $n \rightarrow 0$ limit. The path integral has the form
\bea
\mathcal{Z} &=& \int  \mathcal{D} Q^{ab} (\tau_1, \tau_2) \exp \left( -N \mathcal{S}[Q] \right)\nn
\mathcal{S}[Q] &=&  \frac{3 J^2}{4} \int d \tau_1 d \tau_2 \sum_{ab} \left[ Q^{ab} (\tau_1, \tau_2) \right]^2 + \mathcal{S}_1 [Q]
\label{r1}
\eea
where the functional $\mathcal{S}_1[Q]$ is to obtained by a path integral over the electrons. Near the transition, it turns out to be sufficient to evaluate $\mathcal{S}_1[Q]$ in powers of $J^2$ and $U$ to understand the basic structure of the critical point, as in other theories of the onset of broken symmetry in metals \cite{hertz}. Explicitly the expression is
\bea
\exp\left( -N \mathcal{S}_1[Q] \right) &=& \int \mathcal{D} R^{ab} (\tau_1, \tau_2)  \exp \left( - 
\frac{N t^2}{2} \left| R^{ab} (\tau_1, \tau_2) \right|^2 - N \mathcal{S}_2[Q,R] \right) \nn
\exp\left( -\mathcal{S}_2 [Q,R] \right) &=& \int \mathcal{D} c_{\alpha}^a (\tau) \exp \left\{-
\int d \tau \left[c_\alpha^{a\dagger} (\tau)  \left(\frac{\partial}{\partial \tau}-\mu \right)  c_\alpha^a (\tau) + \frac{U}{2} c^{a\dagger}_\alpha (\tau) c^{a\dagger}_{\beta} (\tau) c_{\beta}^{a} (\tau) c_{\alpha}^{a} (\tau) \right] \right. \nonumber \\
  &+& \left. t^2 \int d\tau d \tau' R^{ab} (\tau,\tau') c_\alpha^{a\dagger} (\tau) c_\alpha^b (\tau')  + \frac{J^2}{2} \int d\tau d \tau' Q^{ab} (\tau, \tau') \vec{S}^a (\tau) \cdot \vec{S}^b (\tau') \right\},
  \label{r2}
\eea
where the second expression generalizes (\ref{Z}).
Analyzing the saddle-point equations of (\ref{r1}) and (\ref{r2}) in the large $N$ limit, it can be verified that we obtain the replica generalizations of the self-consistency conditions in (\ref{self0}) and (\ref{selfcons}).

First, we explicitly solve the saddle-point equations for $R$ at $J^2 = U = 0$. The solution is replica diagonal and depends only on time differences
\beq
R^{ab} (\tau, \tau') = \delta^{ab} R(\tau - \tau')\,.
\eeq
The equation for $R(\tau)$ is easily expressed in frequency space
\beq
R(i \omega) = \frac{1}{i \omega + \mu - t^2 R(i \omega)}\,,
\eeq
and yields the Green's function of a disordered Fermi liquid with the expected semi-circular density of the states
\beq
R(i \omega) = \frac{1}{2t^2} \left( i \omega + \mu - \sqrt{ (i \omega + \mu)^2 - 4 t^2} \right)\,.
\eeq
For $|\mu| < 2 t$, we have a non-zero density of states at the Fermi level, and the low frequency behavior
\beq
R (i \omega) =  \frac{\mu}{2 t^2} - \mbox{sgn} (\omega) \frac{\sqrt{4t^2 - \mu^2}}{2 t^2}\,. \label{r3}
\eeq
This is just the Fourier transform of the electron Green's function in (\ref{Gc}) with $\Delta_f + \Delta_b = 1/2$.

Next we expand (\ref{r2}) in powers of $J^2$ and $U$, and evaluate the path integral over the $c_\alpha^a$.
This results in an expression for $\mathcal{S} [Q]$ as a polynomial in the $Q^{ab} (\tau_1, \tau_2)$ which has the form described in Ref.~\onlinecite{SRO1995}; see also Chapter 22 in Ref.~\onlinecite{qptbook}.
A crucial term in this expansion is a term linear in $Q$ of the form
\bea
-\frac{J^2}{2} \int d\tau d \tau' Q^{ab} (\tau, \tau') \left\langle \vec{S}^a (\tau) \cdot \vec{S}^b (\tau') \right\rangle &\sim&  -\frac{J^2}{2} \int d\tau d \tau' \frac{Q^{aa} (\tau, \tau')}{(\tau - \tau')^2} \nn
& \sim & T J^2 \sum_{\omega_n} Q^{aa} (i \omega_n, -i \omega_n) \left[ r + |\omega_n| \right]\,, \label{r4}
\eea
where $\omega_n = 2 \pi n T$ is a Matsubara frequency. The $1/\tau^2$ term is characteristic of the decay of spin correlations in a disordered Fermi liquid, a direct consequence of the constant density of states at the Fermi level in (\ref{r3}). The frequency dependence in (\ref{r4}) is primarily responsible for the low frequency dynamics of the spin glass order parameter, and the frequency depends of all the other terms in the action for $Q$ can be safely neglected. Near the critical point, it turns out to be sufficient to keep only the following terms in the effective action \cite{SRO1995}
\begin{eqnarray}
\mathcal{S}[Q] &=&  \frac{r}{\kappa}
\int d\tau \,  Q^{aa} (\tau, \tau) - \frac{1}{\pi \kappa} \int d\tau d \tau' \, \frac{Q^{aa} (\tau, \tau')}{(\tau - \tau')^2} 
  \nonumber \\
&&{-}\, \frac{\kappa}{3} \int  d \tau_1 d \tau_2 d \tau_3
\sum_{abc} Q^{ab}( \tau_1 , \tau_2 ) Q^{bc}
( \tau_2 , \tau_3 ) Q^{ca}
( \tau_3 , \tau_1 ) {+}\, \frac{u}{2} \int  d \tau  \left[ 
Q^{aa} ( \tau , \tau) \right]^2 \,.
\label{landau}
\end{eqnarray}
Here we have performed a shift by an unimportant short-time correlator, $Q^{ab} (\tau, \tau') \rightarrow Q^{ab} (\tau, \tau') - C \delta^{ab} \delta(\tau - \tau')$, to eliminate quadratic terms in the action similar to that in (\ref{r1}).

It is now a straightforward matter to solve the saddle point equations of (\ref{landau}). The saddle-point value $Q$ has the following structure in Matsubara frequency space
\beq
Q^{ab} (i \omega_n, i \epsilon_n) = \frac{\delta_{\omega_n,0} \delta_{\epsilon_n,0}}{T^2} \, q^{ab} + \frac{\delta_{\omega_n + \epsilon_n,0}}{T}\,
\delta^{ab}\, Q (i \omega_n)\,. \label{r5}
\eeq
Here $q^{ab}$ is the spin-glass order parameter, which is non-zero only for the parameter $r$ (appearing in (\ref{r4})) smaller than a critical value, $r < r_c$. Focusing first on the Fermi liquid state found for $r>r_c$, it was shown that the second term in (\ref{r5}) has the form
\beq
Q (i \omega_n)  \sim - \sqrt{ r - r_c + |\omega_n| } \quad, \quad r \geq r_c \,.
\eeq
This quantity is just the Fourier transform of $Q (\tau)$
appearing in (\ref{self0}) and (\ref{selfcons}), and so at the critical point $r=r_c$, the spin correlations decay as
\beq
Q (\tau) \sim \frac{1}{|\tau|^{3/2}} \quad , \quad r=r_c \,. \label{r6}
\eeq
Comparing with (\ref{QRpower2}), we see that $\epsilon' = -1/2$. So the present critical point appears distinct from the fixed point $FP_3$ found in Section~\ref{sec:RG}, which can satisfy the self-consistency conditions only for $\epsilon' > 0$.
 
 Finally, let us also recall \cite{RSY95,SRO1995} the nature of the solution for $r< r_c$. Here the spin glass order parameter $q^{ab}$ is non-zero. In the simplest replica-symmetric ansatz, $q^{ab} = q_{EA}$, the Edwards-Anderson order parameter for all $a$ and $b$, and
 \beq
 q_{EA} \sim r_c - r \quad, \quad r < r_c
 \eeq
 At $T>0$, the stable solution has replica symmetry breaking, but the structure of this is very similar to that of the classical spin glass. The replica diagonal term $Q(\tau)$ in (\ref{r5}) is the only one with non-trivial time dependence, and this remains pinned at the $r=r_c$ form in (\ref{r6}) for $r < r_c$.

\bibliography{dqcp}

\end{document}